\documentclass[reqno,10pt,a4paper,dvips]{amsart}

\usepackage{amssymb,times,mathptmx,cite,eucal,psfrag,array,setspace,geometry,color}
\usepackage[dvips]{graphicx}

\numberwithin{equation}{section}

\newcolumntype{C}{>{$}c<{$}} %Defines math mode in tabular (array package)...
\allowdisplaybreaks

\newcommand{\alg}[1]{\mathfrak{#1}}
\newcommand{\group}[1]{\mathsf{#1}}

\newcommand{\Aut}{\group{Aut} \ }
\newcommand{\Out}{\group{Out} \ }

\newcommand{\func}[2]{#1 \left( #2 \right)}
\newcommand{\tfunc}[2]{#1 \bigl( #2 \bigr)}

\newcommand{\brac}[1]{\left( #1 \right)}
\newcommand{\sqbrac}[1]{\left[ #1 \right]}
\newcommand{\set}[1]{\left\{ #1 \right\}}

\newcommand{\abs}[1]{\left| #1 \right|}
\newcommand{\tabs}[1]{\bigl| #1 \bigr|}
\newcommand{\norm}[1]{\left\| #1 \right\|}
\newcommand{\inner}[2]{\left\langle #1 , #2 \right\rangle}

\newcommand{\ZZ}{\mathbb{Z}}
\newcommand{\NN}{\mathbb{N}}
\newcommand{\RR}{\mathbb{R}}
\newcommand{\CC}{\mathbb{C}}

\newcommand{\dd}{\mathrm{d}}
\newcommand{\ii}{\mathfrak{i}}

\newcommand{\eps}{\varepsilon}

\newcommand{\dCox}{\mathsf{h}^{\vee}}

\newcommand{\killing}[2]{\kappa \bigl( #1 , #2 \bigr)}
\newcommand{\invkilling}[2]{\kappa^{-1} \bigl( #1 , #2 \bigr)}
\newcommand{\affine}[1]{\widehat{#1}}

\newcommand{\comm}[2]{\bigl[ #1 , #2 \bigr]}
\newcommand{\acomm}[2]{\bigl\{ #1 , #2 \bigr\}}

\newcommand{\ket}[1]{\bigl\lvert #1 \bigr\rangle}

\newcommand{\bracket}[3]{\bigl\langle #1 \bigr\rvert #2 \bigl\lvert #3 \bigr\rangle} % braket = < | > and bracket = < | | >
\newcommand{\lket}[1]{\left\lvert #1 \right\rangle}

\newcommand{\normord}[1]{{} : #1 : {}} % {} necessary to prevent := or =:

\newcommand{\IndMod}[1]{\mathcal{M}_{#1}}

\newcommand{\AffVerMod}[1]{\affine{\mathcal{V}}_{#1}}
\newcommand{\AffIrrMod}[1]{\affine{\mathcal{L}}_{#1}}
\newcommand{\ExtVerMod}[1]{\mathbb{V}_{#1}}
\newcommand{\ExtIrrMod}[1]{\mathbb{L}_{#1}}

\newcommand{\ch}[2]{\func{\chi_{#1}}{#2}}
\newcommand{\nch}[2]{\func{\widetilde{\chi}_{#1}}{#2}}

\newcommand{\fuse}{\times_{\! f}}

\newcommand{\Jth}[2]{\vartheta_{#1} \bigl( #2 \bigr)}

\newcommand{\qfact}[2]{\left( #1 \right)_{#2}}

\newcommand{\eqnref}[1]{Equation~(\ref{#1})}
\newcommand{\eqnDref}[2]{Equations~(\ref{#1}) and (\ref{#2})}

\newcommand{\secref}[1]{Section~\ref{#1}}
\newcommand{\secsref}[2]{Sections~\ref{#1} -- \ref{#2}}
\newcommand{\appref}[1]{Appendix~\ref{#1}}
\newcommand{\figref}[1]{Figure~\ref{#1}}

\newcommand{\cft}{conformal field theory}
\newcommand{\cfts}{conformal field theories}
\newcommand{\uea}{universal enveloping algebra}
\newcommand{\lcft}{logarithmic conformal field theory}
\newcommand{\lcfts}{logarithmic conformal field theories}
\newcommand{\WZW}{Wess-Zumino-Witten}
\newcommand{\ope}{operator product expansion}
\newcommand{\opes}{operator product expansions}
\newcommand{\hws}{highest weight state}
\newcommand{\hwss}{highest weight states}
\newcommand{\hwm}{highest weight module}
\newcommand{\hwms}{highest weight modules}

\DeclareMathOperator{\id}{id}
\DeclareMathOperator{\vectspan}{span}
\DeclareMathOperator{\tr}{tr}
\DeclareMathOperator{\rank}{rank}
\DeclareMathOperator{\ad}{ad}

\begin{document}

\title{$\func{\affine{\alg{sl}}}{2}_{-1/2}$:  A Case Study}

\author[D Ridout]{David Ridout}

\address[David Ridout]{
Theory Group, DESY \\
Notkestra\ss{}e 85 \\
D-22603, Hamburg, Germany
}

\email{dridout@mail.desy.de}

\thanks{\today}

\begin{abstract}
The construction of the non-logarithmic conformal field theory based on $\func{\affine{\alg{sl}}}{2}_{-1/2}$ is revisited.  Without resorting to free-field methods, the determination of the spectrum and fusion rules is streamlined and the $\beta \gamma$ ghost system is carefully derived as the extended algebra generated by the unique finite-order simple current.  A brief discussion of modular invariance is given and the Verlinde formula is explicitly verified.
\end{abstract}

\maketitle

\onehalfspacing

\section{Introduction} \label{secIntro}

Fractional level \WZW{} models were posited long ago as a tool to construct the non-unitary minimal models.  Their introduction was facilitated by the discovery of Kac and Wakimoto \cite{KacMod88,KacMod88b,KacCla89} of a class of irreducible representations of affine algebras whose (normalised) characters carry a representation of the modular group $\func{\group{SL}}{2 ; \ZZ}$.  These so-called admissible representations include, but are not limited to, the integrable representations from which the rational \WZW{} models are constructed.  The integrable representations necessarily have non-negative integer levels, so the fractional level models must be constructed from non-integrable admissible representations.

Whereas the rational models have a well-known geometric description as non-linear sigma models on compact (simple) group manifolds \cite{WitNon84}, this cannot be generalised to fractional level models.  Indeed, the action defining such a sigma model is ambiguous unless the level is an integer\footnote{This is not necessarily true if one drops the requirement of compactness.  However, investigations of conformally invariant sigma models on non-compact group manifolds have not yet revealed any clear relation to the fractional level models.} \cite{NovMul81}.  Of course an action is not a prerequisite for constructing a \cft{}, especially a non-unitary one, and one can proceed in a purely algebraic manner from the representation theory of the appropriate affine algebra.

At each level there are only finitely many admissible representations.  Indeed, this number is almost always zero, and levels for which this is not the case are sometimes referred to as being admissible themselves.  This finiteness property led to the conjecture that such algebraically-defined fractional level \WZW{} models were also rational \cfts{}.  Indeed, the characters of the admissible representations close under the modular group action and this action is unitary, as in the integer level case.  However, it was quickly realised that the Verlinde formula, which relates the fusion coefficients of the theory to the modular $S$-matrix \cite{VerFus88}, gives negative fusion coefficients in general \cite{KohFus88,BerFoc90,MatFra90}.  Moreover, the matrix representing conjugation, $S^2$, was also observed to contain negative entries.  Even worse, subsequent investigations determining the fusion rules from the decoupling of the null vectors of the representations (in correlation functions) gave different results \cite{AwaFus92,FeiFus94,AndOpe95,PetFus96,FurAdm97}.  Whilst there have been some proposals for how to interpret these negative coefficients \cite{BerFoc90,RamNew93}, this resulted in a general feeling that the fractional level models suffered from an ``intrinsic sickness'' \cite{DiFCon97} and that only their coset theories were well-defined.

All of these efforts were hampered by the seemingly natural assumption that the fusion of admissible modules decomposed into direct sums of admissible modules.  This was pointed out by Gaberdiel \cite{GabFus01}, who studied the ``smallest'' fractional level model corresponding to the affine algebra $\func{\affine{\alg{sl}}}{2}$ at level $\tfrac{-4}{3}$ (smallest in the sense of having the minimal number of admissible representations).  Using a purely algebraic algorithm to compute the fusion rules of the admissible representations \cite{NahQua94,GabInd96}, rather than the Verlinde formula or correlation functions, he was able to show that fusing admissible representations sometimes resulted in reducible but indecomposable representations of the type found in \lcft{}.  Furthermore, he also gave strong evidence that these fusions sometimes produced representations for which the conformal dimensions of the states were \emph{not} bounded below.

This may seem like a textbook definition of ``intrinsic sickness'', but there is a very natural way to understand these unbounded-below representations.  The fusion rules of the rational \WZW{} models respect, in a natural way, the automorphisms of the underlying affine algebra.  It is therefore natural to expect that the fusion rules of the fractional level models will too, and explicit computations completely support this expectation (however, we mention that no proof of this property has yet been advanced).  Whereas these automorphisms transform integrable representations into one another, the same is not true for the admissible representations.  There, one finds that the infinite group of affine algebra automorphisms leads to an infinite number of distinct transformed representations, only a finite number of which have conformal dimensions which are bounded below.

This ruins the hope that fractional level \WZW{} models would be rational \cfts{}, but in a manner which is easy to control.  The inherent irrationality seems to be restricted to these automorphic copies (in modern parlance, the images under \emph{spectral flow}) of the admissible representations.  Of course, there is still the realisation that these models are logarithmic --- work on understanding the nature of the indecomposable representations that arise in these models is still in its infancy.  Nevertheless, this provides a convenient handle with which one can try to understand the true nature of fractional level models.  It is no longer appropriate to regard these models as poorly-defined curiosities.  Rather, it is natural to regard these models as fundamental building blocks for irrational and logarithmic \cfts{}, much as their integer level cousins are for rational theories.

With this in mind, another fractional level model was studied in \cite{LesSU202}, $\func{\affine{\alg{sl}}}{2}$ at level $\tfrac{-1}{2}$.  This model is particularly interesting to field theorists as it has been known for some time (see \cite{GurRel98} for a statement to this effect) that the $\beta \gamma$ system of ghost fields exhibits the same symmetry.  In other words, this fractional level model is equivalent to a free field theory.  Somewhat perversely, the authors of \cite{LesSU202} did not analyse $\func{\affine{\alg{sl}}}{2}_{-1/2}$ using this equivalence, but instead realised it in terms of a different ghost system and a lorentzian boson.  The advantage of this approach was that they were also able to divine the existence of unbounded-below representations in terms of ``multiple-twist'' fields, albeit at a formidable computation cost.  More interestingly, the theory they explored was not logarithmic, in contrast to the $k = \tfrac{-4}{3}$ theory of \cite{GabFus01}.

In this note, we revisit the construction of the $\func{\affine{\alg{sl}}}{2}_{-1/2}$ \cft{}.  Our aim is threefold.  First, we emphasise that this theory is in fact extremely easy to analyse if one abandons free-field constructions.  We do so here in an expository fashion which makes it clear how to generalise to other admissible levels.  Indeed, $k = \tfrac{-1}{2}$ is the first of an infinite series of admissible levels $k = \tfrac{1}{2} \brac{2m-1}$ ($m \in \NN$) which give rise to non-\lcfts{}.  We expect that all other admissible levels give rise to logarithmic theories.  Our second aim is to make precise the relation between the algebra $\func{\affine{\alg{sl}}}{2}_{-1/2}$ and the ghost algebra considered in \cite{LesSU202}.  This provides an excellent example of the extended algebra formalism of \cite{RidSU206,RidMin07}, in which all the subtleties uncovered there are present.  Our last aim is to point out that there is nothing mysterious or ``sick'' about the modular properties of this theory.  The partition functions, conjugation matrices and Verlinde formula all work \emph{exactly} as expected.

The organisation is as follows.  After first introducing our notations and conventions (\secref{secAlg}), we derive the structure of the irreducible vacuum module in \secref{secVacMod}.  It is worthwhile seeing explicitly in at least one case that admissibility just means that the corresponding Verma module has the same ``braided'' singular vector structure as the integrable modules.  This gives us the ``null-vector constraints'' on the other representations of the theory, thence the other admissible \hwms{} (\secref{secAdmiss}).  Character formulae for all these are derived.

We then proceed to the computation of the fusion rules of the theory (\secref{secFusion}).  This involves considering the purely algebraic algorithm of Nahm \cite{NahQua94} and Gaberdiel-Kausch \cite{GabInd96}.  Whilst this algorithm is computationally intensive, we note that by making two very plausible assumptions, we do not actually have to perform any computations and can proceed using only logical consequences of the algorithm.  First, we assume that the irreducible vacuum module acts as the fusion identity.  We could of course prove this easily using the fusion algorithm, but prefer to note that this assumption is consistent unless we uncover a logarithmic partner state to the vacuum (which we do not).  Logic alone then allows us to compute the fusion rules of the admissible modules.  In particular, we prove that certain fusions of admissible modules lead to modules whose conformal dimensions are unbounded below.  The second assumption then allows us to identify these modules.  This is the assumption that the fusion rules respect the \emph{spectral flow} automorphisms of $\func{\affine{\alg{sl}}}{2}$.

This then gives us a complete infinite spectrum of irreducible modules which closes under fusion.  In \secref{secChar}, we determine the full set of characters of the theory, noting that they are not all independent as one might expect from rational theories.  Instead, there are only four linearly independent characters.  We argue, following \cite{LesSU202}, that a module is determined by its character \emph{and} a prescription of how to expand it.  The latter is implicit in rational theories, but the presence of unbounded-below modules (and non-integrable modules in general) forces its explicit acknowledgement here.  The consequent lack of a bijection between the modules and the characters therefore leads us to introduce a Grothendieck ring of characters.

\secsref{secGhost}{secGhostReps} are devoted to a detailed study of the extended algebra of the $\func{\affine{\alg{sl}}}{2}_{-1/2}$ algebra, which is the $\beta \gamma$ ghost system.  In \secref{secGhost}, we show that the (chiral) primary fields defining this extension cannot be taken to be mutually bosonic with the affine currents, and that associativity of the operator product algebra forces the introduction of an additional operator into the theory.  The bosonic ghost fields are defined, but they are not mutually bosonic with respect to the affine currents either.  At issue here is the definition of the adjoint, an integral part of any symmetry algebra.  In \secref{secGhost2}, we change the adjoint and repeat the analysis of the previous section finding satisfying simplifications --- all fields are mutually bosonic and the operator product algebra is associative without need of additional operators.

We then briefly discuss (\secref{secGhostReps}) the representation theory of this extended algebra, remarking upon the consistency of the monodromy charge and the lifted extended algebra spectral flow.  The Verma modules of the extended algebra are verified to be irreducible --- in this sense the $\beta \gamma$ ghost system may be said to be free --- and fermionic character formulae for them and their $\func{\affine{\alg{sl}}}{2}_{-1/2}$ counterparts are derived.  These formulae give simple expressions for the string functions of all the modules of the theory.

Finally, we conclude by reconsidering the modular properties of the $\func{\affine{\alg{sl}}}{2}_{-1/2}$ theory in \secref{secMod}.  We derive the $S$ and $T$-matrices of the theory, verify that they are symmetric and unitary, and write down a complete set of modular invariants.  Moreover, we check that $S^2$ represents conjugation and the Verlinde formula recovers the fusion coefficients in the Grothendieck ring of characters.

There are also two appendices, the second of which (\appref{appTheta}) is just a summary of our notations and conventions for Jacobi theta functions.  The first, \appref{appSpecFlow}, gives a detailed description of the spectral flow automorphisms as affine Weyl group translations (by elements of the coroot lattice) and affine outer automorphisms (as translations by elements of the dual root lattice).  We are not aware of a comprehensive discussion of this viewpoint in the literature, so we hope that this will be of independent use in the future.

\section{Algebraic Preliminaries} \label{secAlg}

Let $\func{\alg{sl}}{2}$ be the complex Lie algebra spanned by three generators $E$, $H$ and $F$ subject to the commutation relations
\begin{equation}
\comm{H}{E} = 2 E, \qquad \comm{E}{F} = H \qquad \text{and} \qquad \comm{H}{F} = -2 F.
\end{equation}
We define the Killing form to be the trace of the product in the defining (fundamental) two-dimensional representation (equivalently, $1/4$ of the trace of the product in the adjoint representation).  This gives
\begin{equation} \label{eqnKilling}
\killing{H}{H} = 2 \qquad \text{and} \qquad \killing{E}{F} = 1,
\end{equation}
with all other combinations vanishing.  The affine Kac-Moody algebra $\func{\affine{\alg{sl}}}{2}$ is then the vector space
\begin{equation}
\func{\alg{sl}}{2} \otimes \CC \sqbrac{t ; t^{-1}} \oplus \vectspan_{\CC} \set{K , L_0}
\end{equation}
equipped with the commutation relations
\begin{subequations} \label{eqnCommRels}
\begin{gather}
\comm{J^a_m}{J^b_n} = \comm{J^a}{J^b}_{m+n} + m \killing{J^a}{J^b} \delta_{m+n,0} K, \qquad \comm{J^a_m}{K} = 0, \\
\comm{L_0}{J^a_m} = -m J^a_m \qquad \text{and} \qquad \comm{L_0}{K} = 0.
\end{gather}
\end{subequations}
Here, $J^a_m$ denotes $J^a \otimes t^m$, where $J^a$ can represent $H$, $E$ or $F$.  We are generally interested in representations of $\func{\affine{\alg{sl}}}{2}$ on which the central element $K$ acts as $k$ times the identity, for some common scalar $k$ called the level.  In what follows, we will be principally interested in the case where $k = \tfrac{-1}{2}$.

As is well known, the \uea{} of $\func{\affine{\alg{sl}}}{2}$ contains a subalgebra isomorphic to the (\uea{} of the) Virasoro algebra (when $k \neq -2$).  This is the Sugawara construction.  Here, the Virasoro elements are realised as quadratic elements normally ordered in the standard way:
\begin{equation} \label{eqnDefVir}
L_n = \frac{1}{2 \brac{k+2}} \sum_{r \in \ZZ} \normord{\frac{1}{2} H_r H_{n-r} + E_r F_{n-r} + F_r E_{n-r}}.
\end{equation}
As usual, we will identify $L_0 \in \func{\affine{\alg{sl}}}{2}$ with the quadratic element $L_0$ constructed in \eqnref{eqnDefVir}.  The central charge defined by the Sugawara construction is $c = 3k / \brac{k+2}$.

We define a triangular decomposition of $\func{\affine{\alg{sl}}}{2}$ as follows:  The span of $H_0$, $K$ and $L_0$ defines the Cartan subalgebra, the raising operators are $E_{n-1}$, $H_n$ and $F_n$ for $n \geqslant 1$, and the adjoint is defined by
\begin{equation} \label{eqnDefAdj}
E_n^{\dag} = F_{-n}, \qquad H_n^{\dag} = H_{-n}, \qquad K^{\dag} = K \qquad \text{and} \qquad L_n^{\dag} = L_{-n}.
\end{equation}
We can now talk about \hwss{} and Verma modules.  It is easy to check from \eqnref{eqnDefVir} that an affine \hws{} with $\func{\alg{sl}}{2}$-weight ($H_0$-eigenvalue) $\lambda$ has conformal dimension ($L_0$-eigenvalue)
\begin{equation}
h_{\lambda} = \frac{\lambda \brac{\lambda + 2}}{4 \brac{k+2}}.
\end{equation}
The $\func{\alg{sl}}{2}$-weight $\lambda$, conformal dimension $h$ and the level $k$ completely determine an $\func{\affine{\alg{sl}}}{2}$-weight $\affine{\lambda} = \brac{\lambda , k , h}$.  As the level is given and the conformal dimension of a \hws{} is determined by its $\func{\alg{sl}}{2}$-weight, it follows that an $\func{\affine{\alg{sl}}}{2}$-Verma module is characterised solely by the latter.  We therefore denote Verma modules by $\AffVerMod{\lambda}$.

The fundamental question to ask about Verma modules concerns their reducibility.  If a Verma module contains a proper submodule, then this submodule is generated by singular vectors, non-trivial descendant \hwss{}.  Quotienting $\AffVerMod{\lambda}$ by its maximal proper submodule gives the corresponding irreducible module $\AffIrrMod{\lambda}$.  To find singular vectors, we can use the fact that Verma modules come equipped with a unique (up to normalisation) invariant inner product defined by the adjoint (\ref{eqnDefAdj}), the Shapovalov form.  With respect to this form, the (non-trivial) singular vectors and their descendants are all null, meaning that their norm is zero.  The presence of such null states can be detected by computing the determinant of the Shapovalov form in each affine weight space.

Happily, there is an explicit form for this determinant, given by the Kac-Kazhdan formula \cite{KacStr79}:  The Shapovalov determinant of $\AffVerMod{\lambda}$ in the weight space $\brac{\lambda - \mu , k , h_{\lambda} + m}$ is
\begin{multline} \label{eqnKacKazhdan}
\func{{\textstyle \det_{\lambda}}}{\mu , m} = \prod_{\ell=1}^{\infty} \Biggl\{ \brac{\lambda + 1 - \ell}^{\func{P}{-\mu + 2 \ell , m}} \prod_{n=1}^{\infty} \bigl( \lambda + 1 + n \brac{k+2} - \ell \bigr)^{\func{P}{-\mu + 2 \ell , m - n \ell}} \Biggr. \\
\Biggl. \cdot \bigl( -\lambda - 1 + n \brac{k+2} - \ell \bigr)^{\func{P}{-\mu - 2 \ell , m - n \ell}} \bigl( n \brac{k+2} \bigr)^{\func{P}{-\mu , m - n \ell}} \Biggr\},
\end{multline}
where $\func{P}{\mu , m}$ denotes the multiplicity with which the weight $\brac{\mu , 0 , m}$ appears in the module $\AffVerMod{0}$ (this is independent of $k$).  The presence of a singular vector in $\AffVerMod{\lambda}$ is signalled by the vanishing of one of the factors appearing in this formula and the vanishing of the arguments of the function $P$ occurring in the corresponding exponent (non-vanishing arguments of this $P$ in general correspond to descendants of the singular vector).  We will refer to weights which admit a singular vector as singular weights.

\section{Vacuum Module Structure} \label{secVacMod}

We now specialise to $k = \tfrac{-1}{2}$, with the aim of constructing a \cft{}.  This theory will therefore have central charge $c = -1$.  The first step is to determine a vacuum module.  By definition, the vacuum $\ket{0}$ is an $\func{\affine{\alg{sl}}}{2}$-\hws{} which is also annihilated by all the zero-modes, in particular by $H_0$ and $F_0$.  The vacuum module is therefore a quotient module of $\AffVerMod{0}$.  We can analyse these quotients by determining the singular vector structure of the Verma module, and to do this we use the Kac-Kazhdan formula (\ref{eqnKacKazhdan}).

Setting $\lambda = 0$ in this formula, we see that the determinant vanishes when
\begin{equation}
\ell = 1, \qquad \ell = \frac{3n}{2} + 1 \qquad \text{or} \qquad \ell = \frac{3n}{2} - 1 \qquad \text{($n \in 2 \ZZ_+$).}
\end{equation}
In the first case, the arguments of $P$ in the corresponding exponent vanish if $\mu = 2 \ell = 2$ and $m = 0$, indicating that the singular vector has weight $\brac{-2,\tfrac{-1}{2},0}$.  This is clearly the singular vector $F_0 \ket{0}$ which is set to zero by definition.  The other two cases are more interesting and the weights of the corresponding singular vectors are found to be
\begin{equation} \label{eqnSW1}
\brac{-6m-2,\frac{-1}{2},2m \brac{3m+1}} \qquad \text{and} \qquad \brac{6m-2,\frac{-1}{2},2m \brac{3m-1}} \qquad \text{($m = \frac{n}{2} \in \ZZ_+$)},
\end{equation}
respectively.  The first few singular weights are therefore
\begin{equation}
\brac{4,\frac{-1}{2},4}, \quad \brac{-8,\frac{-1}{2},8}, \quad \brac{10,\frac{-1}{2},20}, \quad \brac{-14,\frac{-1}{2},28}, \quad \ldots
\end{equation}
These weights do not determine the singular vector itself, but it can be shown that every weight space of a Verma module admits at most one singular vector.

Unfortunately, these are not the only singular weights of $\AffVerMod{0}$.  We also have to check for singular vectors which are descended from those we have already found.  In other words, we should check the submodules which the known singular vectors generate for further singular vectors.  Repeating the above Kac-Kazhdan analysis for the submodule generated by the singular weight $\brac{-2,\tfrac{-1}{2},0}$, we find further singular weights of the form
\begin{equation} \label{eqnSW2}
\brac{-6m,\frac{-1}{2},2m \brac{3m-1}} \qquad \text{and} \qquad \brac{6m,\frac{-1}{2},2m \brac{3m+1}} \qquad \text{($m \in \ZZ_+$)}.
\end{equation}
These describe two series of singular vectors which are completely disjoint from those found above.  The first few weights are
\begin{equation}
\brac{-6,\frac{-1}{2},4}, \quad \brac{6,\frac{-1}{2},8}, \quad \brac{-12,\frac{-1}{2},20}, \quad \brac{12,\frac{-1}{2},28}, \quad \ldots
\end{equation}

\begin{figure}
\begin{center}
\includegraphics[width=13cm]{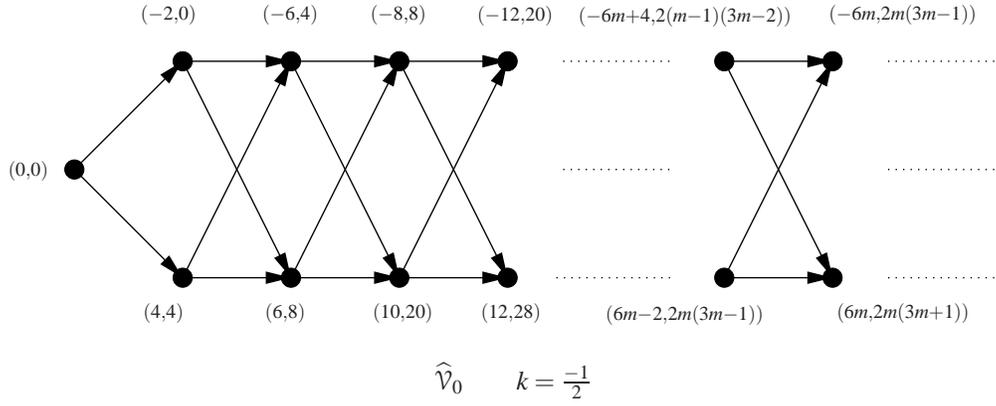}
\caption{The singular vector structure of the vacuum Verma module at level $\tfrac{-1}{2}$.  Each singular vector is labelled by its $\func{\alg{sl}}{2}$-weight and conformal dimension (respectively).} \label{figVacVerMod}
\end{center}
\end{figure}

However, the weight $\brac{-2,\tfrac{-1}{2},0}$ and those given in (\ref{eqnSW1}) and (\ref{eqnSW2}) exhaust the singular weights of $\AffVerMod{0}$.  This is not hard to check explicitly:  For example, the singular weights descended from that of weight $\brac{-6m-2,\frac{-1}{2},2m \brac{3m+1}}$ of (\ref{eqnSW1}) all have the form
\begin{equation}
\begin{split}
\brac{-6 \brac{m'-m},\frac{-1}{2},2 \brac{m'-m} \brac{3 \brac{m'-m} - 1}} \qquad \text{($m' > 2m$)} \\
\text{or} \qquad \brac{6 \brac{m'+m},\frac{-1}{2},2 \brac{m'+m} \brac{3 \brac{m'+m} + 1}} \qquad \text{($m' \in \ZZ_+$),}
\end{split}
\end{equation}
which are both of the form given in (\ref{eqnSW2}).  It follows now that the singular vector structure of $\AffVerMod{0}$ is as shown in \figref{figVacVerMod}.  Note the braiding pattern familiar from the integrable modules (and the Virasoro algebra).

\psfrag{00}[][]{$\scriptstyle \brac{0,0}$}
\psfrag{-20}[][]{$\scriptstyle \brac{-2,0}$}
\psfrag{44}[][]{$\scriptstyle \brac{4,4}$}
\psfrag{-64}[][]{$\scriptstyle \brac{-6,4}$}
\psfrag{68}[][]{$\scriptstyle \brac{6,8}$}
\psfrag{-88}[][]{$\scriptstyle \brac{-8,8}$}
\psfrag{1020}[][]{$\scriptstyle \brac{10,20}$}
\psfrag{-1220}[][]{$\scriptstyle \brac{-12,20}$}
\psfrag{1228}[][]{$\scriptstyle \brac{12,28}$}
\psfrag{aaaa}[][]{$\scriptstyle \brac{-6m+4,2 \brac{m-1} \brac{3m-2}}$}
\psfrag{cccc}[][]{$\scriptstyle \brac{6m-2,2m \brac{3m-1}}$}
\psfrag{bbbb}[][]{$\scriptstyle \brac{-6m,2m \brac{3m-1}}$}
\psfrag{dddd}[][]{$\scriptstyle \brac{6m,2m \brac{3m+1}}$}
\psfrag{V}[][]{$\AffVerMod{0} \qquad k = \frac{-1}{2}$}

It follows that the descendant singular vectors of $\AffVerMod{0}$ are generated by the two singular vectors of weights $\brac{-2,\tfrac{-1}{2},0}$ and $\brac{4,\tfrac{-1}{2},4}$.  The former is the vector $F_0 \ket{0}$ which we have already set to zero, so we see that there are only two possible choices for the vacuum module.  Either we set the dimension $4$ singular vector to zero, or we do not.  We choose to set this singular vector to zero, thereby taking the vacuum module to be the irreducible quotient $\AffIrrMod{0}$.  The alternative, in which this singular vector is not set to zero, will undoubtedly lead to a logarithmic \cft{} \cite{GabAlg03} (assuming it can be defined), which we do not want to consider here\footnote{We remark that a logarithmic \cft{} with $\func{\affine{\alg{sl}}}{2}_{-1/2}$ symmetry was proposed in \cite{LesLog04}, based on free field constructions.  We do not expect that keeping the singular vector in the vacuum module will lead to this theory.  We intend to return to a detailed discussion of how the theory discussed here can be extended to something similar to that of \cite{LesLog04} in a future publication.}.

The character for the vacuum Verma module is easily computed from the standard Poincar\'{e}-Birkhoff-Witt basis and has the form
\begin{equation} \label{eqnChV0}
\ch{\AffVerMod{0}}{z ; q} = {\textstyle \tr_{\AffVerMod{0}}} z^{H_0} q^{L_0} = \frac{1}{\displaystyle \prod_{i=1}^{\infty} \brac{1 - z^{-2} q^{i-1}} \brac{1 - q^i} \brac{1 - z^2 q^i}} = \frac{1}{\displaystyle \sum_{n \in \ZZ} \brac{-1}^n z^{2n} q^{n \brac{n+1} / 2}},
\end{equation}
where we have used Jacobi's triple product identity, \eqnref{eqnJacobiTriple}, in the last step.  It now follows from the embedding pattern of \figref{figVacVerMod} that the character of the (irreducible) vacuum module takes the form
\begin{align} \label{eqnCh0}
\ch{\AffIrrMod{0}}{z ; q} &= \sqbrac{1 - \sum_{n=1}^{\infty} \brac{z^{6n-2} q^{2n \brac{3n-1}} + z^{-6n+4} q^{2 \brac{n-1} \brac{3n-2}}} + \sum_{n=1}^{\infty} \brac{z^{6n} q^{2n \brac{3n+1}} + z^{-6n} q^{2n \brac{3n-1}}}} \ch{\AffVerMod{0}}{z ; q} \notag \\
&= \frac{\displaystyle \sum_{n \in \ZZ} \brac{z^{-6n} - z^{6n-2}} q^{2n \brac{3n-1}}}{\displaystyle \prod_{i=1}^{\infty} \brac{1 - z^{-2} q^{i-1}} \brac{1 - q^i} \brac{1 - z^2 q^i}} = \frac{\displaystyle \sum_{n \in \ZZ} \brac{z^{-6n} - z^{6n-2}} q^{2n \brac{3n-1}}}{\displaystyle \sum_{n \in \ZZ} \brac{-1}^n z^{2n} q^{n \brac{n+1} / 2}}.
\end{align}

\section{Admissible Representations} \label{secAdmiss}

Now that we have a vacuum module, we can ask if it constrains the spectrum of the theory.  Since the vacuum module has a vanishing singular vector at $\func{\alg{sl}}{2}$-weight $4$ and grade $4$, the answer is ``yes'' (setting $F_0 \ket{0}$ to zero does not affect the spectrum as it is a part of the definition of the vacuum).  To derive the constraints, we need the explicit form of this singular vector.  There exist semi-explicit formulae for such singular vectors in the literature \cite{MalSin86,BauFus93,MatPri99}, but it is not hard to compute it directly in this case.  It turns out to be
\begin{equation} \label{eqnVacSV}
\brac{156 E_{-3} E_{-1} - 71 E_{-2}^2 + 44 E_{-2} H_{-1} E_{-1} - 52 H_{-2} E_{-1}^2 - 16 F_{-1} E_{-1}^3 - 4 H_{-1}^2 E_{-1}^2} \ket{0} = 0,
\end{equation}
up to normalisation.  By the state-field correspondence of \cft{}, this vanishing singular vector gives rise to a vanishing chiral field whose modes must therefore annihilate any physical state \cite{FeiAnn92}.  These are the constraints we seek.

Instead of considering this singular vector itself, it is convenient to consider its $\func{\alg{sl}}{2}$-weight $0$ descendant obtained by acting with $F_0^2$.  This descendant field is (up to normalisation)
\begin{multline}
\Lambda = 64 \normord{EEFF} - 16 \normord{EHHF} + 136 \normord{EH \partial F} - 128 \normord{E \partial HF} + 12 \normord{E \partial^2 F} - 8 \normord{HHHH} \\
- 200 \normord{\partial EHF} +108 \normord{\partial E \partial F} + 8  \normord{\partial HHH} - 38 \normord{\partial H \partial H} - 156 \normord{\partial^2 EF} + 24 \normord{\partial^2 HH} - \partial^3 H.
\end{multline}
Let $\ket{\lambda}$ be a \hws{} with $\func{\alg{sl}}{2}$-weight $\lambda$.  Since the modes of $\Lambda$ must annihilate any physical state,
\begin{equation}
0 = \Lambda_0 \ket{\lambda} = \brac{-8 H_0^4 - 8 H_0^3 - 38 H_0^2 + 48 H_0^2 + 6 H_0} \ket{\lambda} = -2 \lambda \brac{\lambda - 1} \brac{2 \lambda + 1} \brac{2 \lambda + 3} \ket{\lambda},
\end{equation}
implying that $\lambda = 0, 1, \tfrac{-1}{2}, \tfrac{-3}{2}$.  These are the only allowed \hwss{} of the theory.  Their conformal dimensions are $0$, $\tfrac{1}{2}$, $\tfrac{-1}{8}$ and $\tfrac{-1}{8}$, respectively.

\begin{figure}
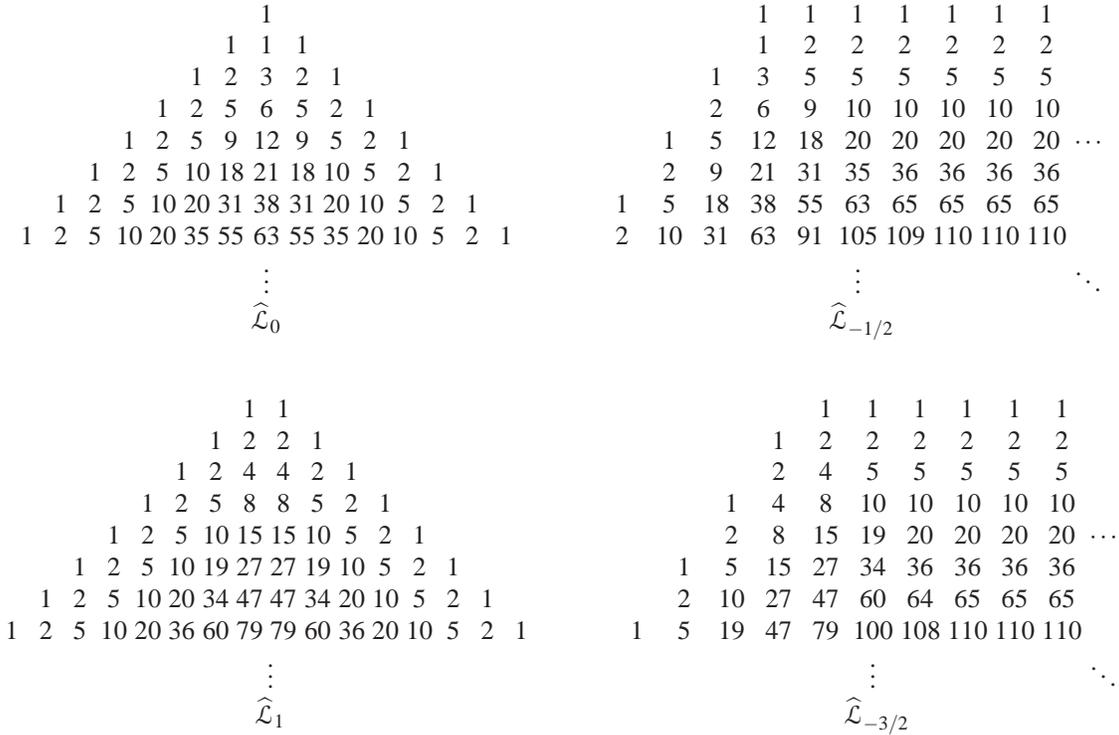

\begin{center}
\begin{tabular}{*{15}{@{}C@{\hspace{1mm}}}@{}C@{\hspace{10mm}}*{11}{@{}C@{\hspace{1mm}}}}
  &  &  &  &  &  &  & 1&  &  &  &  &  &  &  & &   &   &   &  1&  1&  1&  1&  1&  1&  1&   \\
  &  &  &  &  &  & 1& 1& 1&  &  &  &  &  &  & &   &   &   &  1&  2&  2&  2&  2&  2&  2&   \\
  &  &  &  &  & 1& 2& 3& 2& 1&  &  &  &  &  & &   &   &  1&  3&  5&  5&  5&  5&  5&  5&   \\
  &  &  &  & 1& 2& 5& 6& 5& 2& 1&  &  &  &  & &   &   &  2&  6&  9& 10& 10& 10& 10& 10&   \\
  &  &  & 1& 2& 5& 9&12& 9& 5& 2& 1&  &  &  & &   &  1&  5& 12& 18& 20& 20& 20& 20& 20&\cdots\\
  &  & 1& 2& 5&10&18&21&18&10& 5& 2& 1&  &  & &   &  2&  9& 21& 31& 35& 36& 36& 36& 36&   \\
  & 1& 2& 5&10&20&31&38&31&20&10& 5& 2& 1&  & &  1&  5& 18& 38& 55& 63& 65& 65& 65& 65&   \\
 1& 2& 5&10&20&35&55&63&55&35&20&10& 5& 2& 1& &  2& 10& 31& 63& 91&105&109&110&110&110&   \\
\phantom{55}&\phantom{55}&\phantom{55}&\phantom{55}&\phantom{55}&\phantom{55}&\phantom{55}&\vdots&\phantom{55}&\phantom{55}&\phantom{55}&\phantom{55}&\phantom{55}&\phantom{55}&\phantom{55}& &\phantom{155}&\phantom{155}&\phantom{155}&\phantom{155}&\phantom{155}&\vdots&\phantom{155}&\phantom{155}&\phantom{155}&\phantom{155}&\ddots\\
  &  &  &  &  &  &  &\AffIrrMod{0}&  &  &  &  &  &  &  & &  &  &  &  &\multicolumn{3}{C}{\AffIrrMod{-1/2}}&  &  &  &  \\
\end{tabular}
\\[7mm]
\begin{tabular}{*{16}{@{}C@{\hspace{1mm}}}@{}C@{\hspace{10mm}}*{11}{@{}C@{\hspace{1mm}}}}
  &  &  &  &  &  &  & 1& 1&  &  &  &  &  &  &  & &   &   &   &   &  1&  1&  1&  1&  1&  1&   \\
  &  &  &  &  &  & 1& 2& 2& 1&  &  &  &  &  &  & &   &   &   &  1&  2&  2&  2&  2&  2&  2&   \\
  &  &  &  &  & 1& 2& 4& 4& 2& 1&  &  &  &  &  & &   &   &   &  2&  4&  5&  5&  5&  5&  5&   \\
  &  &  &  & 1& 2& 5& 8& 8& 5& 2& 1&  &  &  &  & &   &   &  1&  4&  8& 10& 10& 10& 10& 10&   \\
  &  &  & 1& 2& 5&10&15&15&10& 5& 2& 1&  &  &  & &   &   &  2&  8& 15& 19& 20& 20& 20& 20&\cdots\\
  &  & 1& 2& 5&10&19&27&27&19&10& 5& 2& 1&  &  & &   &  1&  5& 15& 27& 34& 36& 36& 36& 36&   \\
  & 1& 2& 5&10&20&34&47&47&34&20&10& 5& 2& 1&  & &   &  2& 10& 27& 47& 60& 64& 65& 65& 65&   \\
 1& 2& 5&10&20&36&60&79&79&60&36&20&10& 5& 2& 1& &  1&  5& 19& 47& 79&100&108&110&110&110&   \\
\phantom{55}&\phantom{55}&\phantom{55}&\phantom{55}&\phantom{55}&\phantom{55}&\phantom{55}&\multicolumn{2}{C}{\vdots}&\phantom{55}&\phantom{55}&\phantom{55}&\phantom{55}&\phantom{55}&\phantom{55}&\phantom{55}& &\phantom{155}&\phantom{155}&\phantom{155}&\phantom{155}&\phantom{155}&\vdots&\phantom{155}&\phantom{155}&\phantom{155}&\phantom{155}&\ddots\\
  &  &  &  &  &  &  &\multicolumn{2}{C}{\AffIrrMod{1}}&  &  &  &  &  &  &  & &  &  &  &  &\multicolumn{3}{C}{\AffIrrMod{-3/2}}&  &  &  &  \\
\end{tabular}
\caption{The multiplicities of the weights of the admissible representations of $\func{\affine{\alg{sl}}}{2}_{-1/2}$.  In these pictures, the $\func{\alg{sl}}{2}$-weight increases from right to left (in multiples of $2$) and the conformal dimension increases from top to bottom (in multiples of $1$).} \label{figAdmissReps}
\end{center}
\end{figure}

Now that we know the possible \hwss{}, we can ask about the possible \hwms{}.  For example, repeating the analysis of \secref{secVacMod} shows that the Verma module $\AffVerMod{1}$ has singular vectors of weights $\brac{-3,\frac{-1}{2},\frac{1}{2}}$,
\begin{equation}
\brac{\pm 6m-3,\frac{-1}{2},\frac{1}{2} + 2m \brac{3m \mp 2}} \qquad \text{and} \qquad \brac{\pm 6m+1,\frac{-1}{2},\frac{1}{2} + 2m \brac{3m \pm 2}} \qquad \text{($m \in \ZZ_+$).}
\end{equation}
The embedding pattern is again of braided type and the two generating singular vectors are those with weights
\begin{equation} \label{eqnV1SVs}
\brac{-3,\frac{-1}{2},\frac{1}{2}} \qquad \text{and} \qquad \brac{3,\frac{-1}{2},\frac{5}{2}}.
\end{equation}
The $\func{\alg{sl}}{2}$-weights of these singular vectors do not belong to the allowed set, hence we can conclude that all descendant singular vectors vanish in the physical module.  It follows that the \hws{} with $\lambda = 1$ generates the irreducible module $\AffIrrMod{1}$.  Its character is
\begin{equation} \label{eqnCh1}
\ch{\AffIrrMod{1}}{z ; q} = z q^{1/2} \ \frac{\displaystyle \sum_{n \in \ZZ} \brac{z^{-6n} - z^{6n-4}} q^{2n \brac{3n-2}}}{\displaystyle \sum_{n \in \ZZ} \brac{-1}^n z^{2n} q^{n \brac{n+1} / 2}}.
\end{equation}
Similarly, one finds that the modules corresponding to $\lambda = \tfrac{-1}{2}$ and $\tfrac{-3}{2}$ are also irreducible with characters
\begin{align}
\ch{\AffIrrMod{-1/2}}{z ; q} &= z^{-1/2} q^{-1/8} \ \frac{\displaystyle \sum_{n \in \ZZ} \brac{z^{6n} q^{n \brac{6n+1}} - z^{-6n+2} q^{\brac{2n-1} \brac{3n-1}}}}{\displaystyle \sum_{n \in \ZZ} \brac{-1}^n z^{2n} q^{n \brac{n+1} / 2}} \label{eqnCh-1/2} \\
\text{and} \qquad \ch{\AffIrrMod{-3/2}}{z ; q} &= z^{-3/2} q^{-1/8} \ \frac{\displaystyle \sum_{n \in \ZZ} \brac{z^{6n} q^{n \brac{6n-1}} - z^{-6n+4} q^{\brac{2n-1} \brac{3n-2}}}}{\displaystyle \sum_{n \in \ZZ} \brac{-1}^n z^{2n} q^{n \brac{n+1} / 2}}, \label{eqnCh-3/2}
\end{align}
respectively.  These are the \emph{admissible} \hwms{} of Kac and Wakimoto \cite{KacMod88b}.  We illustrate them in \figref{figAdmissReps}.

\section{Fusion and the Spectrum} \label{secFusion}

We turn now to the derivation of the fusion rules of the admissible modules.  Normally, we could investigate this by computing $3$-point correlation functions of the primary fields.  However, doing this requires making a number of non-trivial assumptions.  In particular, we must assume that every field has a conjugate so that the matrix of $2$-point functions is non-degenerate.  Moreover, we would also be implicitly assuming that we have already identified every field of the theory.  Since there are no candidates within the admissible representations for the conjugate fields to the dimension $\tfrac{-1}{8}$ primaries, we must conclude that there are further fields to discover.  But if we admit to not knowing the field content of the theory, then it follows that we cannot be sure that the $2$-point functions we will use in our fusion computations are non-degenerate.  For example, it seems reasonable to declare that the vacuum is self-conjugate, so that the $2$-point function of the identity is constant.  However, if subsequent fusion computations revealed that the vacuum had a logarithmic partner state, then it would follow from general principles \cite{FloBit03} that the conjugate field to the identity would be this logarithmic partner (and the $2$-point function of the identity would vanish identically), contradicting our original declaration.

For this reason, we will be careful and compute fusion using a purely algebraic method that makes no reference to correlation functions nor non-degeneracy.  This is the algorithm of Nahm and Gaberdiel-Kausch \cite{NahQua94,GabInd96}.  Happily, the situation here is sufficiently simple that we will not have to make any explicit computations with this algorithm; we will be able to proceed with a few logical consequences which are easy to state (and hopefully understand).

In general, this algorithm constructs a representation $\Delta$ of the symmetry algebra on the fusion product of two modules $\IndMod{1}$ and $\IndMod{2}$.  Decomposing this representation gives the fusion rule $\IndMod{1} \fuse \IndMod{2}$.  In practice, one only constructs this representation to a chosen finite grade $g$ --- all the ``deeper'' structure of the fused module is thrown away.  The idea is to choose $g$ large enough that one obtains as much information as is required.  The representation $\Delta$ is constructed within the \emph{working space}, which for affine symmetry algebras consists of the tensor product of the zero-grade subspace of $\IndMod{1}$ and the subspace of $\IndMod{2}$ consisting of elements with grade at most $g$.  The working space is then reduced by removing the so-called \emph{spurious states} which reflect the vanishing of certain singular vectors of $\IndMod{1}$ and $\IndMod{2}$.  This is achieved by employing three master equations \cite[Eqs.~2.2--2.4]{GabInd96} iteratively on expressions formed from these singular vectors.  The fusion representation space is thereby constructed (to grade $g$) when all spurious states are removed.  The master equations then define the action of the symmetry algebra (that is $\Delta$) upon what remains.

It is extremely important to note that we need to assume that the conformal dimensions of the states composing each module are bounded from below\footnote{In fact, one can sometimes bypass this requirement \cite{GabFus01}, but it adds significantly to the complexity of the computations.}.  Then, we can define the \emph{grade} of an arbitrary state in this module to be the difference between the dimension of the state and the minimal dimension.  It should be clear that we require this bounded-below property for the modules we are fusing \emph{and} the modules we generate via fusion.  Indeed, when we say that the fusion representation space is constructed to grade $g$, we mean that upon decomposition, the structure of each component module is determined to grade $g$ in the above sense.

We will not need to enter into the details of this algorithm.  Computing to grade $0$ turns out to suffice for our purposes, and so we will only determine the action of the $\func{\alg{sl}}{2}$-subalgebra spanned by the zero-modes $E_0$, $H_0$ and $F_0$.  The master equations give this action as
\begin{equation}
\func{\Delta}{J_0} = J_0 \otimes \id + \id \otimes J_0 \qquad \text{($J = E, H, F$),}
\end{equation}
which is identical to the tensor product $\func{\alg{sl}}{2}$-action (on $\func{\alg{sl}}{2}$-modules).  The fusion of $\func{\affine{\alg{sl}}}{2}$-modules to grade $0$ therefore only differs from the tensor product of the corresponding grade $0$ $\func{\alg{sl}}{2}$-modules if there are non-trivial spurious states.  We point out that if we find a spurious state, then acting upon it with any $\func{\Delta}{J_0}$ must give another spurious state.  The spurious states therefore form a representation of the zero-mode $\func{\alg{sl}}{2}$-subalgebra.

Consider first fusing the vacuum module $\AffIrrMod{0}$ with some other module $\IndMod{}$.  We require only that the zero-grade states of $\IndMod{}$ form an irreducible $\func{\alg{sl}}{2}$-module.  The working space is then the tensor product of the trivial $\func{\alg{sl}}{2}$-module with this irreducible $\func{\alg{sl}}{2}$-module.  There are vanishing singular vectors in at least one of the $\func{\affine{\alg{sl}}}{2}$-modules, so there could be spurious states.  But, the working space is isomorphic to a single irreducible $\func{\alg{sl}}{2}$-module, so the existence of spurious states would mean that the fusion product is empty!  The possibilities are therefore that fusing a module with the vacuum gives the module back again or nothing.

Suppose therefore that $\AffIrrMod{0} \fuse \IndMod{}$ is empty for some module $\IndMod{}$ in our theory.  As $\IndMod{}$ must have a conjugate representation $\IndMod{}^+$ in the theory, $\IndMod{} \fuse \IndMod{}^+ = \AffIrrMod{0} + \ldots$.  By hypothesis, the result of fusing the left hand side of this rule with $\AffIrrMod{0}$ is empty, hence $\AffIrrMod{0} \fuse \AffIrrMod{0}$ must also be empty.  But this implies that the vacuum is a null state, which requires the existence of a logarithmic partner (as we noted above).  We may therefore proceed under the assumption that $\AffIrrMod{0} \fuse \IndMod{}$ is not empty for any module in our theory --- this will only be invalidated if we find that it leads to a logarithmic partner to the vacuum.  As we will see, we do not find this outcome, hence it is consistent to insist that the irreducible vacuum module $\AffIrrMod{0}$ acts as the fusion identity on every module in the theory\footnote{Of course, we can explicitly show that $\AffIrrMod{0} \fuse \IndMod{} = \IndMod{}$ for each of our admissible modules using the Nahm-Gaberdiel-Kausch algorithm.  But the above argument is much more elementary, and has the additional advantage of drawing attention to the subtleties possible when one does find logarithmic structure.  It does assume the existence of conjugates, however this is physically necessary in all (quasirational) theories, even logarithmic ones (with a suitable interpretation) --- fields without a conjugate decouple within correlation functions.}.  Note that it follows from this that the vacuum module is self-conjugate.

A more interesting computation is to determine the fusion of the module $\AffIrrMod{1}$ with itself.  Computing to grade $0$ again, we may regard the working space as the tensor product of the fundamental representation of $\func{\alg{sl}}{2}$ with itself.  This decomposes as the direct sum of the trivial and adjoint representations, so the working space contains a $\func{\alg{sl}}{2}$-\hws{} of weight $2$.  In the absence of any spurious states, this would imply that the fused module contains a $\func{\affine{\alg{sl}}}{2}$-\hws{} of weight $\brac{2, \tfrac{-1}{2}, \tfrac{4}{3}}$.  But this is forbidden by the vacuum singular vector (\secref{secAdmiss}), so the weight $2$ \hws{} must be spurious.  It then follows that the entire adjoint representation must also be spurious, so we are left with the trivial $\func{\alg{sl}}{2}$-representation.  If this were also spurious, then the fusion product would be empty.  The requirement of a conjugate for $\AffIrrMod{1}$ would then force $\AffIrrMod{0} \fuse \AffIrrMod{1}$ to be empty, contradicting the fact that $\AffIrrMod{0}$ is the fusion identity.  The $\func{\affine{\alg{sl}}}{2}$-module corresponding to the trivial $\func{\alg{sl}}{2}$-representation is clearly the vacuum module\footnote{Matthias Gaberdiel points out that this assumes that the result of the fusion is a module whose conformal dimensions are bounded below.  I believe that this assumption is warranted because of the finite-dimensionality of the working space, but I have no proof of this at present.  In any case, the conclusion of the above argument has been confirmed by explicitly calculating the fusion structure to grade $1$ using the full Nahm-Gaberdiel-Kausch algorithm.}, so we have derived the following fusion rule:
\begin{equation} \label{eqnFR1x1}
\AffIrrMod{1} \fuse \AffIrrMod{1} = \AffIrrMod{0}.
\end{equation}
The module $\AffIrrMod{1}$ is therefore also self-conjugate.

We can continue in a similar fashion to determine the fusion rules of the other admissible modules.  In particular, we obtain
\begin{equation} \label{eqnFR1xAdmiss}
\AffIrrMod{1} \fuse \AffIrrMod{-1/2} = \AffIrrMod{-3/2} \qquad \text{and} \qquad \AffIrrMod{1} \fuse \AffIrrMod{-3/2} = \AffIrrMod{-1/2}
\end{equation}
without fuss.  The rest of the fusion rules are more delicate to analyse however.  For example, considering $\AffIrrMod{-1/2} \fuse \AffIrrMod{-1/2}$ to grade $0$ as above, the working space decomposes as an $\func{\alg{sl}}{2}$-representation into an infinite direct sum\footnote{We know that this is a direct sum because these $\func{\alg{sl}}{2}$-representations are unitary with respect to the $\func{\alg{sl}}{2 ; \RR}$ adjoint $J^{\ddag} = -J$, $J = E,H,F$.} of irreducibles whose highest weights are $-1, -3, -5, \ldots$.  Proceeding as above, we would conclude that none of these highest weights are allowed, hence that all states are spurious and the fusion product is empty.

But as with $\AffIrrMod{1}$, insisting on a conjugate for $\AffIrrMod{-1/2}$, even if we have not yet identified it, again leads to a contradiction.  The loophole is in trusting that a non-spurious $\func{\alg{sl}}{2}$-\hws{} corresponds to an $\func{\affine{\alg{sl}}}{2}$-\hws{}.  We could trust this correspondence in the previous examples because the Nahm-Gaberdiel-Kausch algorithm gives us, grade by grade, the affine structure of the fused module.  However, this algorithm does not make sense if the conformal dimensions of the states of the fused module are not bounded from below (recall that in this situation, the concept of grade is not defined).  Before, this boundedness property was guaranteed because we only had finitely many irreducible representations of the zero-mode $\func{\alg{sl}}{2}$-subalgebra.  In the case at hand however, there are infinitely many such representations, so the conformal dimension need not be bounded from below.  Indeed, we cannot even compute the conformal dimensions of the $\func{\alg{sl}}{2}$-\hwss{} without assuming something about the fused module structure.  The correct conclusion to draw from our analysis is that either $\AffIrrMod{-1/2} \fuse \AffIrrMod{-1/2}$ is empty, which leads to a contradiction, or that it gives a module whose states have arbitrarily negative conformal dimension.

In fact, it is not too difficult to justify directly that the second option is what actually occurs.  To do this, we make use of the automorphisms of our symmetry algebra, in particular the \emph{spectral flow} automorphisms (described in detail in \appref{appSpecFlow}).  For $\func{\affine{\alg{sl}}}{2}$, the spectral flow is freely generated by a single automorphism $\gamma$ which may be taken to act by (see \eqnref{eqnSpecFlowA1})
\begin{subequations} \label{eqnSF}
\begin{gather}
\func{\gamma}{E_n} = E_{n-1}, \qquad \func{\gamma}{H_n} = H_n - \delta_{n,0} K, \qquad \func{\gamma}{F_n} = F_{n+1}, \\
\func{\gamma}{K} = K \qquad \text{and} \qquad \func{\gamma}{L_0} = L_0 - \frac{1}{2} H_0 + \frac{1}{4} K.
\end{gather}
\end{subequations}
We consider the induced action of $\gamma$ on the vacuum.  Specifically, we determine the $\func{\alg{sl}}{2}$-weight and conformal dimension of $\tfunc{\gamma}{\ket{0}}$:
\begin{align}
H_0 \func{\gamma}{\ket{0}} &= \func{\gamma}{\func{\gamma^{-1}}{H_0} \ket{0}} = \func{\gamma}{\brac{H_0 + K} \ket{0}} = \frac{-1}{2} \func{\gamma}{\ket{0}}, \\
L_0 \func{\gamma}{\ket{0}} &= \func{\gamma}{\func{\gamma^{-1}}{L_0} \ket{0}} = \func{\gamma}{\Bigl( L_0 + \frac{1}{2} H_0 + \frac{1}{4} K \Bigr) \ket{0}} = \frac{-1}{8} \func{\gamma}{\ket{0}}.
\end{align}
Similarly, one can check that $\func{\gamma}{\ket{0}}$ is a \hws{}.  This therefore suggests that
\begin{equation} \label{eqnMapI}
\func{\gamma}{\AffIrrMod{0}} = \AffIrrMod{-1/2}.
\end{equation}
This is to be interpreted as $\gamma \circ \pi_0 \circ \gamma^{-1} = \pi_{-1/2}$, where $\pi_{\lambda}$ denotes the representation (map) of $\func{\affine{\alg{sl}}}{2}$ on $\AffIrrMod{\lambda}$.  Note that here $\gamma^{-1}$ is acting as an automorphism on $\func{\affine{\alg{sl}}}{2}$, whereas $\gamma$ denotes the induced isomorphism of vector spaces acting on the states (as in $\tfunc{\gamma}{\ket{0}}$ above).

We can prove (\ref{eqnMapI}) in many ways.  First, we can note that what we have proven is the corresponding equality of Verma modules.  We should therefore explicitly check that the expressions for the two vanishing singular vectors of each module are mapped to zero by $\gamma$ and $\gamma^{-1}$ (as appropriate).  A second proof involves verifying that the characters satisfy
\begin{align} \label{eqnCh1stSF}
\ch{\AffIrrMod{-1/2}}{z ; q} &= \ch{\func{\gamma}{\AffIrrMod{0}}}{z ; q} = \sum_{\text{basis } \func{\gamma}{\lket{\psi}}} z^{\lambda_{\func{\gamma}{\lket{\psi}}}} q^{\Delta_{\func{\gamma}{\lket{\psi}}}} \notag \\
&= \sum_{\text{basis } \lket{\psi}} z^{\lambda_{\lket{\psi}} + k} q^{\Delta_{\lket{\psi}} + \frac{1}{2} \lambda_{\lket{\psi}} + \frac{1}{4} k} = z^{-1/2} q^{-1/8} \ch{\AffIrrMod{0}}{z q^{1/2} ; q},
\end{align}
where $\lambda_{\lket{\psi}}$ and $\Delta_{\lket{\psi}}$ denote the $\func{\alg{sl}}{2}$-weight and conformal dimension of $\ket{\psi}$ (respectively).  This can be done in a straight-forward fashion using the character formulae given in \eqnDref{eqnCh0}{eqnCh-1/2}.  However, it is far more elegant to simply observe that twisting a representation by an algebra automorphism clearly preserves irreducibility.  To this third proof, we add the practical method of looking at the picture of $\AffIrrMod{0}$ in \figref{figAdmissReps}, turning one's head $45^{\circ}$ to the right, and comparing with the picture of $\AffIrrMod{-1/2}$ there.  In this way, we observe that the multiplicities of the appropriate weight spaces precisely match.  This actually constitutes a rigorous proof in itself because the pictures in \figref{figAdmissReps} show the multiplicities to sufficiently deep grades (in general, both pictures must show that the generating singular vectors vanish).

We can similarly study the spectral flow of $\AffIrrMod{1}$.  Proceeding as above, we compute that the image of the \hws{} $\ket{1}$ under $\gamma$ has $\func{\alg{sl}}{2}$-weight $\tfrac{1}{2}$ and conformal dimension $\tfrac{7}{8}$.  However, it is not a \hws:
\begin{equation}
F_1 \func{\gamma}{\ket{1}} = \func{\gamma}{\func{\gamma^{-1}}{F_1} \ket{1}} = \func{\gamma}{F_0 \ket{1}} \neq 0.
\end{equation}
Instead, it is the image of $F_0 \ket{1}$ which becomes the \hws{} of the flowed module $\tfunc{\gamma}{\AffIrrMod{1}}$.  This can be checked explicitly, but is most easily seen from \figref{figAdmissReps}.  Since $\func{\gamma}{F_0 \ket{1}}$ has $\func{\alg{sl}}{2}$-weight $\tfrac{-1}{2}$ and conformal dimension $\tfrac{-1}{8}$, it now follows that
\begin{equation}
\func{\gamma}{\AffIrrMod{1}} = \AffIrrMod{-3/2}.
\end{equation}

We can apply our new-found knowledge regarding the action of the spectral flow automorphisms to the computation of fusion rules.  This relies upon the principle that the fusion rules respect these automorphisms in the following manner\footnote{This can only apply when the automorphisms acting commute.  It is not clear what should replace this principle in general.}:
\begin{equation}
\IndMod{} \fuse \IndMod{}' = \IndMod{}'' \qquad \Rightarrow \qquad \func{\Omega}{\IndMod{}} \fuse \func{\Omega'}{\IndMod{}'} = \func{\Omega \Omega'}{\IndMod{}''}.
\end{equation}
Here, $\Omega$ and $\Omega'$ are automorphisms.  This principle is well-known from studies of rational \cfts{} with Lie algebra symmetries.  Despite its natural appearance, we are not aware of any formal general proof.  It has however been checked explicitly in many non-trivial cases (see \cite{GabFus01} in particular).  For example, we can determine $\AffIrrMod{1} \fuse \AffIrrMod{-1/2}$ by applying $\id \fuse \gamma$ (in hopefully obvious notation) to $\AffIrrMod{1} \fuse \AffIrrMod{0}$.  The result reproduces the first fusion rule of (\ref{eqnFR1xAdmiss}).

More importantly, we can apply this principle to compute the fusion of $\AffIrrMod{-1/2}$ with itself.  The result is therefore that this gives the module $\tfunc{\gamma^2}{\AffIrrMod{0}} = \tfunc{\gamma}{\AffIrrMod{-1/2}}$ (and indeed we see that the fusion is not empty).  This module is not one of the admissibles that we have considered.  Indeed, it is not even a \hwm{}, as can be seen by looking at the picture of $\AffIrrMod{-1/2}$ in \figref{figAdmissReps} and turning one's head $45^{\circ}$ to the right.  This is perhaps physically distasteful, but is an unavoidable feature of the theory.  We remark that the conformal dimensions of the states in this module with a given $\func{\alg{sl}}{2}$-weight \emph{are} bounded below.  It follows from this that operator products of the corresponding fields may still be expanded as a Laurent series (with poles of finite order).  The standard field-theoretic machinery of \cft{} may therefore be carried across to these modules without difficulty.

\begin{figure}
\begin{center}
\includegraphics[width=15cm]{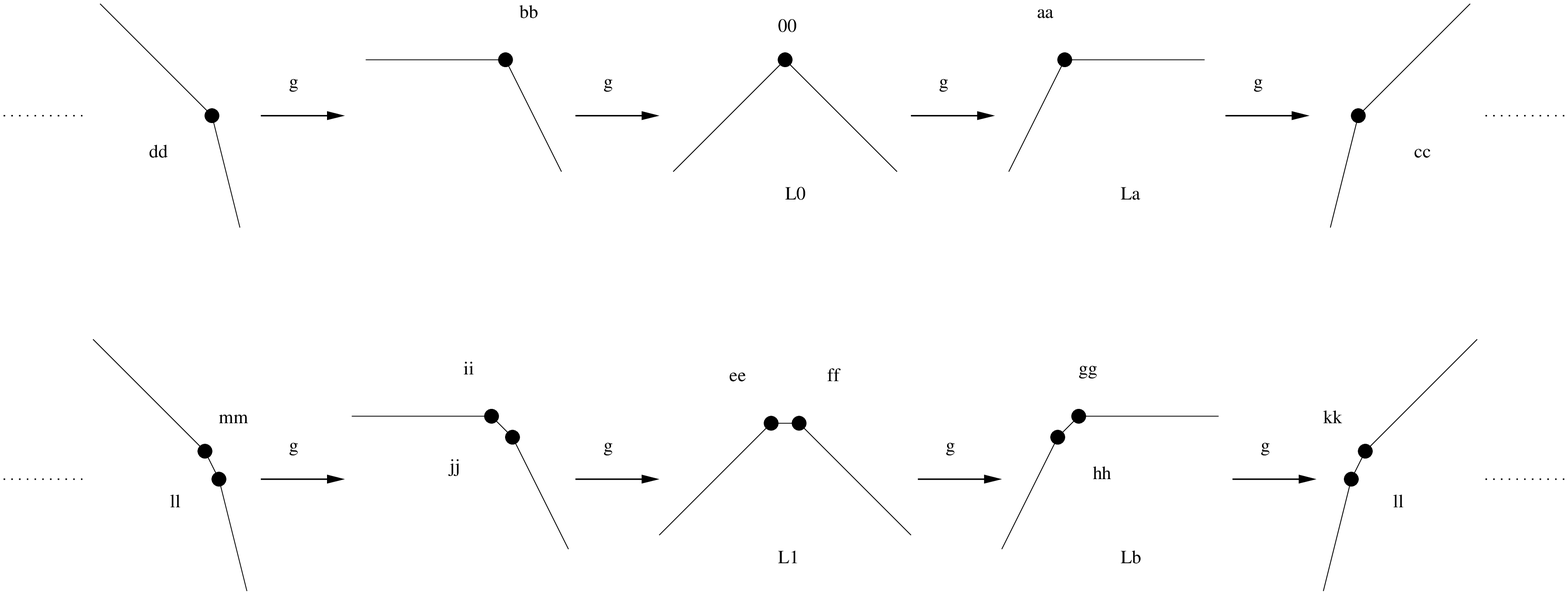}
\caption{Depictions of the modules appearing in the spectrum and the action of the spectral flow automorphism $\gamma$.  Each ``corner state'' is labelled by its $\func{\alg{sl}}{2}$-weight and conformal dimension (in that order).} \label{figSpecFlow}
\end{center}
\end{figure}

It is now trivial to determine the remaining fusion rules of the admissible modules:
\begin{equation}
\AffIrrMod{-1/2} \fuse \AffIrrMod{-1/2} = \AffIrrMod{-3/2} \fuse \AffIrrMod{-3/2} = \func{\gamma^2}{\AffIrrMod{0}} \qquad \text{and} \qquad \AffIrrMod{-1/2} \fuse \AffIrrMod{-3/2} = \func{\gamma^2}{\AffIrrMod{1}}.
\end{equation}
Moreover, we now see that the spectrum contains the modules $\tfunc{\gamma^{\ell}}{\AffIrrMod{0}}$ and $\tfunc{\gamma^{\ell}}{\AffIrrMod{1}}$ for all $\ell$.  Extending this to $\ell$ negative also makes sense, and is in fact necessary for physical consistency.  Otherwise (for example), $\AffIrrMod{-1/2}$ would have no conjugate within the spectrum, so correlation functions of its fields with any other fields would vanish, leading to the effective decoupling (and removal) of $\AffIrrMod{-1/2}$ from the theory.  The conjugate of $\AffIrrMod{-1/2} = \tfunc{\gamma}{\AffIrrMod{0}}$ is of course $\tfunc{\gamma^{-1}}{\AffIrrMod{0}}$.  This is not a \hwm{} --- one pictures it by looking at $\AffIrrMod{0}$ in \figref{figAdmissReps} and turning one's head $45^{\circ}$ to the left --- as its zero-grade states form a \emph{lowest} weight representation of $\func{\alg{sl}}{2}$.  We indicate this module (and other flowed modules) schematically in \figref{figSpecFlow}.  Note that even the Weyl group of $\func{\alg{sl}}{2}$ does not preserve the modules in the spectrum:  The non-trivial reflection induces a (grade-preserving) map between $\tfunc{\gamma^{\ell}}{\AffIrrMod{\lambda}}$ and $\tfunc{\gamma^{-\ell}}{\AffIrrMod{\lambda}}$ (for $\lambda = 0 , 1$).  Of course, this map is nothing but conjugation (as usual for $\func{\affine{\alg{sl}}}{2}$ theories).

\psfrag{L0}[][]{$\AffIrrMod{0}$}
\psfrag{L1}[][]{$\AffIrrMod{1}$}
\psfrag{La}[][]{$\AffIrrMod{-1/2}$}
\psfrag{Lb}[][]{$\AffIrrMod{-3/2}$}
\psfrag{g}[][]{$\gamma$}
\psfrag{00}[][]{$\scriptstyle \brac{0,0}$}
\psfrag{aa}[][]{$\scriptstyle \brac{\tfrac{-1}{2},\tfrac{-1}{8}}$}
\psfrag{bb}[][]{$\scriptstyle \brac{\tfrac{1}{2},\tfrac{-1}{8}}$}
\psfrag{cc}[][]{$\scriptstyle \brac{-1,\tfrac{-1}{2}}$}
\psfrag{dd}[][]{$\scriptstyle \brac{1,\tfrac{-1}{2}}$}
\psfrag{ee}[][]{$\scriptstyle \brac{1,\tfrac{1}{2}}$}
\psfrag{ff}[][]{$\scriptstyle \brac{-1,\tfrac{1}{2}}$}
\psfrag{gg}[][]{$\scriptstyle \brac{\tfrac{-3}{2},\tfrac{-1}{8}}$}
\psfrag{hh}[][]{$\scriptstyle \brac{\tfrac{1}{2},\tfrac{7}{8}}$}
\psfrag{ii}[][]{$\scriptstyle \brac{\tfrac{3}{2},\tfrac{-1}{8}}$}
\psfrag{jj}[][]{$\scriptstyle \brac{\tfrac{-1}{2},\tfrac{7}{8}}$}
\psfrag{kk}[][]{$\scriptstyle \brac{-2,-1}$}
\psfrag{ll}[][]{$\scriptstyle \brac{0,1}$}
\psfrag{mm}[][]{$\scriptstyle \brac{2,-1}$}

We have therefore shown that the the spectrum of our $\func{\affine{\alg{sl}}}{2}_{-1/2}$ theory consists of two infinite series of modules, $\tfunc{\gamma^{\ell}}{\AffIrrMod{0}}$ and $\tfunc{\gamma^{\ell}}{\AffIrrMod{1}}$ ($\ell \in \ZZ$).  The fusion rules may be summarised by
\begin{equation} \label{eqnFR}
\tfunc{\gamma^{\ell}}{\AffIrrMod{\lambda}} \fuse \tfunc{\gamma^{m}}{\AffIrrMod{\mu}} = \tfunc{\gamma^{\ell + m}}{\AffIrrMod{\lambda + \mu}},
\end{equation}
where $\lambda$ and $\mu$ take value $0$ or $1$ and their sum is taken \emph{modulo} $2$.  It should be clear that all the modules in the spectrum are mutually distinct (there are no module isomorphisms between them).

\section{Characters and Modular Invariants} \label{secChar}

Consider now the characters of the modules comprising our theory.  We have already determined the characters of $\AffIrrMod{0}$ and $\AffIrrMod{1}$ and their images under $\gamma$, and it is easy to use the explicit spectral flow action to determine expressions for those which remain.  But let us first take this opportunity to rewrite the known characters in a more standard form \cite{KacMod88b}.  Recall the explicit form of the vacuum character, given in \eqnref{eqnCh0}.  We split the denominator (as an infinite sum over $n$) into sums over $n$ even and $n$ odd.  Completing the square in the $q$-exponents of both the numerator and denominator then gives
\begin{equation} \label{eqnChar0}
\ch{\AffIrrMod{0}}{z ; q} = q^{-1/24} \frac{\displaystyle \sum_{r \in \ZZ + 1/6} z^{6r} q^{6r^2} - \sum_{r \in \ZZ - 1/6} z^{6r} q^{6r^2}}{\displaystyle \sum_{r \in \ZZ + 1/4} z^{4r} q^{2r^2} - \sum_{r \in \ZZ - 1/4} z^{4r} q^{2r^2}}.
\end{equation}
The reader will no doubt recognise that the numerator and denominator are differences of classical theta functions \cite{KacInf90}.  The factor $q^{-1/24}$ is the standard modular anomaly $q^{c/24}$.  A similar massaging of \eqnref{eqnCh1} gives
\begin{equation} \label{eqnChar1}
\ch{\AffIrrMod{1}}{z ; q} = q^{-1/24} \frac{\displaystyle \sum_{r \in \ZZ + 1/3} z^{6r} q^{6r^2} - \sum_{r \in \ZZ - 1/3} z^{6r} q^{6r^2}}{\displaystyle \sum_{r \in \ZZ + 1/4} z^{4r} q^{2r^2} - \sum_{r \in \ZZ - 1/4} z^{4r} q^{2r^2}}.
\end{equation}
Apply now the spectral flow automorphism $\gamma^{\ell}$.  Generalising \eqnref{eqnCh1stSF}, we quickly derive that
\begin{equation} \label{eqnCharSF}
\ch{\tfunc{\gamma^{\ell}}{\AffIrrMod{\lambda}}}{z ; q} = z^{-\ell / 2} q^{-\ell^2 / 8} \ch{\AffIrrMod{\lambda}}{z q^{\ell / 2} ; q}.
\end{equation}
From \eqnDref{eqnChar0}{eqnChar1} we therefore obtain
\begin{align}
\ch{\tfunc{\gamma^{\ell}}{\AffIrrMod{0}}}{z ; q} &= q^{-1/24} \frac{\displaystyle \sum_{r \in \ZZ + \brac{3 \ell + 2} / 12} z^{6r} q^{6r^2} - \sum_{r \in \ZZ + \brac{3 \ell - 2} / 12} z^{6r} q^{6r^2}}{\displaystyle \sum_{r \in \ZZ + \brac{2 \ell + 1} / 4} z^{4r} q^{2r^2} - \sum_{r \in \ZZ + \brac{2 \ell - 1} / 4} z^{4r} q^{2r^2}} \label{eqnCharSF0} \\
\text{and} \qquad \ch{\tfunc{\gamma^{\ell}}{\AffIrrMod{1}}}{z ; q} &= q^{-1/24} \frac{\displaystyle \sum_{r \in \ZZ + \brac{3 \ell + 4} / 12} z^{6r} q^{6r^2} - \sum_{r \in \ZZ + \brac{3 \ell - 4} / 12} z^{6r} q^{6r^2}}{\displaystyle \sum_{r \in \ZZ + \brac{2 \ell + 1} / 4} z^{4r} q^{2r^2} - \sum_{r \in \ZZ + \brac{2 \ell - 1} / 4} z^{4r} q^{2r^2}}. \label{eqnCharSF1}
\end{align}

This appears to provide a satisfying answer to the determination of the characters of our theory.  However, it is easy to check from \eqnDref{eqnCharSF0}{eqnCharSF1} that the common denominator is antiperiodic under $\ell \rightarrow \ell + 1$, hence periodic under $\ell \rightarrow \ell + 2$.  Moreover, the numerators of the spectrally-flowed characters of $\AffIrrMod{0}$ and $\AffIrrMod{1}$ are interchanged (with an additional factor of $-1$) under $\ell \rightarrow \ell + 2$, and are thus periodic under $\ell \rightarrow \ell + 4$.  It therefore follows that these expressions for the spectrally-flowed characters are periodic in $\ell$ with period $4$, and that there are only four linearly independent characters (there are actually eight distinct characters, but four are just the negatives of the other four).  We can take these to be the characters of the admissible \hwms{} $\AffIrrMod{0}$, $\AffIrrMod{1}$, $\AffIrrMod{-1/2} = \tfunc{\gamma}{\AffIrrMod{0}}$ and $\AffIrrMod{-3/2} = \tfunc{\gamma}{\AffIrrMod{1}}$.  The spectral flow action on the characters may then be summarised as
\begin{equation} \label{eqnLinIndChars}
\begin{split}
\cdots \overset{\gamma}{\longrightarrow} -\chi_{\AffIrrMod{1}} \overset{\gamma}{\longrightarrow} -\chi_{\AffIrrMod{-3/2}} \overset{\gamma}{\longrightarrow} \chi_{\AffIrrMod{0}} \overset{\gamma}{\longrightarrow} \chi_{\AffIrrMod{-1/2}} \overset{\gamma}{\longrightarrow} -\chi_{\AffIrrMod{1}} \overset{\gamma}{\longrightarrow} \cdots \\
\cdots \overset{\gamma}{\longrightarrow} -\chi_{\AffIrrMod{0}} \overset{\gamma}{\longrightarrow} -\chi_{\AffIrrMod{-1/2}} \overset{\gamma}{\longrightarrow} \chi_{\AffIrrMod{1}} \overset{\gamma}{\longrightarrow} \chi_{\AffIrrMod{-3/2}} \overset{\gamma}{\longrightarrow} -\chi_{\AffIrrMod{0}} \overset{\gamma}{\longrightarrow} \cdots
\end{split}
\end{equation}
This seems to contradict the fact that the corresponding modules are all distinct.  There are no isomorphisms between the spectrally-flowed modules, but nevertheless there is an infinite degeneracy of the characters.

A resolution to this seeming contradiction was proposed in \cite{LesSU202}, where it was noted that one has to pay close attention to the regions of convergence of such character formulae.  The problem is very much related to the more transparent example of $\func{\alg{sl}}{2}$ characters.  Here, a highest weight Verma module with highest weight $\lambda$ has character
\begin{equation}
z^{\lambda} + z^{\lambda - 2} + z^{\lambda - 4} + \ldots = \frac{z^{\lambda}}{1 - z^{-2}} \qquad \text{($\abs{z} > 1$).}
\end{equation}
Similarly, a lowest weight Verma module with lowest weight $\lambda + 2$ has character
\begin{equation}
z^{\lambda + 2} + z^{\lambda + 4} + z^{\lambda + 6} + \ldots = \frac{z^{\lambda + 2}}{1 - z^2} \qquad \text{($\abs{z} < 1$).}
\end{equation}
Formally, these characters give the same function (up to a conspicuous factor of $-1$), but the notion that the modules are (almost) the same is patently absurd.  The point is that in general the physical module is determined by its character \emph{and} the given region of convergence.  It is the latter which dictates the formal expansion, here in powers of $z^2$ or $z^{-2}$.  Note that finite-dimensional $\func{\alg{sl}}{2}$-modules have characters that are polynomial in $z$ and $z^{-1}$, hence converge when $\abs{z} = 1$ (indeed, everywhere).

The $\func{\affine{\alg{sl}}}{2}_{-1/2}$ character formulae we have derived have to be understood in a similar way.  Specifically, the infinite sums appearing in the numerators and denominators of these formulae are easily checked to converge for all $z \in \CC$, provided that $\abs{q} < 1$.  However, the common denominator of these expressions vanishes whenever $z^2 = q^i$ ($i \in \ZZ$).  This is obvious from its product form (displayed in \eqnref{eqnChV0} for example), but it is also useful to check this from the above sum form:
\begin{align}
\sum_{r \in \ZZ + \tfrac{2 \ell + 1}{4}} q^{2r^2 + 2ir} - \sum_{r \in \ZZ + \tfrac{2 \ell - 1}{4}} q^{2r^2 + 2ir} &= q^{-i^2 / 2} \sqbrac{\sum_{s \in \ZZ + \tfrac{\ell + i}{2} + \tfrac{1}{4}} q^{2s^2} - \sum_{s \in \ZZ + \tfrac{\ell + i}{2} - \tfrac{1}{4}} q^{2s^2}} \notag \\
&= q^{-i^2 / 2} \sqbrac{\sum_{s \in \ZZ - \tfrac{\ell + i}{2} - \tfrac{1}{4}} q^{2s^2} - \sum_{s \in \ZZ + \tfrac{\ell + i}{2} - \tfrac{1}{4}} q^{2s^2}} = 0,
\end{align}
as $\ell + i \in \ZZ$.  The character formulae will therefore have poles at $z^2 = q^i$ ($i \in \ZZ$) unless the zeroes of the denominator are cancelled by zeroes in the numerator (this is what happens in the integrable module case).  But, analysing the numerators of \eqnDref{eqnCharSF0}{eqnCharSF1} as above, we find that zeroes occur only at $z^2 = q^{\ell + i}$ with $\ell + i \in 2 \ZZ$.  It follows that the character formulae we have given for the modules $\tfunc{\gamma^{\ell}}{\AffIrrMod{0}}$ and $\tfunc{\gamma^{\ell}}{\AffIrrMod{1}}$ have poles at $z^2 = q^i$ for all $i \in 2 \ZZ - 1 - \ell$.

We have argued above with the example of $\func{\alg{sl}}{2}$ that the relationship between a character and the module it is supposed to describe is determined by the region in which the character is to be expanded.  It is now clear how this applies to the present case.  The characters given for $\AffIrrMod{0}$ and $\AffIrrMod{1}$ should be expanded on the annulus $\abs{q}^{1/2} < \abs{z} < \abs{q}^{-1/2}$.  Note that as $\abs{q} < 1$, this covers the case $\abs{z} = 1$.  Accordingly, when $q$-expanding \eqnDref{eqnChar0}{eqnChar1}, the coefficients simplify to give (Laurent) polynomials in $z$ as the constituent $\func{\alg{sl}}{2}$-modules are all finite-dimensional.  The spectral flow action now implies that the appropriate region on which to properly expand the characters (\ref{eqnCharSF0}) and (\ref{eqnCharSF1}) with $\ell \neq 0$ is the annulus
\begin{equation} \label{eqnAnnulus}
\abs{q}^{\brac{-\ell + 1}/2} < \abs{z} < \abs{q}^{\brac{-\ell - 1}/2}.
\end{equation}
Note that when $\ell > 0$, we have $\abs{z} > 1$, appropriate for highest weight $\func{\alg{sl}}{2}$-modules, and for $\ell < 0$, we have $\abs{z} < 1$, appropriate for lowest weight $\func{\alg{sl}}{2}$-modules.  This accords with the pictures we have drawn in \figref{figSpecFlow}.

For $\ell = \pm 1$, $q$-expanding the character gives coefficients which are rational functions of $z$.  These coefficients may then be expanded for either $\abs{z} > 1$ or $\abs{z} < 1$, as appropriate, in order to recover the correct weight multiplicities of the module.  However, for $\abs{\ell} > 1$, this procedure fails.  For example, $q$-expanding \eqnref{eqnCharSF0} with $\ell = 2$ gives polynomial coefficients in $z$ because this character coincides with that of $\AffIrrMod{1}$ up to an overall factor of $-1$ (\eqnref{eqnLinIndChars}).  The expansion annulus (\ref{eqnAnnulus}) for $\ell = 2$ is disjoint from that for $\ell = 0$, so a na\"{\i}ve $q$-expansion\footnote{Here, we mean an expansion in which the powers of $q$ are bounded from below, such as one obtains from computer algebra packages.} is no longer appropriate.  Indeed, for this module the conformal dimension is not bounded below, so the correct expansion would have to include arbitrarily negative powers of $q$ as well as the usual positive powers.  To obtain such an expansion, we would have to change variables to $u = \brac{z q^{\ell / 2}}^{-1}$, effectively undoing the spectral flow, then $q$-expand and change $u$ back again.

It is clear therefore that the explicit expressions we have given for the modules $\tfunc{\gamma^{\ell}}{\AffIrrMod{0}}$ and $\tfunc{\gamma^{\ell}}{\AffIrrMod{1}}$ with $\abs{\ell} > 1$ are not actually particularly useful.  The point however is that any other expression we might cook up will be equivalent to these because the classical theta functions from which they are constructed are entire in the $z$-plane (when $\abs{q} < 1$).  The conclusion is that by expressing the characters in terms of these functions, instead of as formal power series, we lose the equivalence between modules and characters.  We have an infinite collection of distinct modules, but only four linearly independent characters.

More precisely, the $\ZZ$-linear map which assigns to each module in the fusion ring its character is not one-to-one.  It is not hard to check that the kernel of this map is generated by the modules $\tfunc{\gamma^{\ell \pm 1}}{\AffIrrMod{0}} \oplus \tfunc{\gamma^{\ell \mp 1}}{\AffIrrMod{1}}$ and that these modules are closed under fusion.  It follows that we can consistently define fusion at the level of characters.  We call the resulting ring over $\ZZ$ the \emph{Grothendieck ring} of characters\footnote{We should mention that the notion of Grothendieck ring which we have defined here is not quite the same as that used in \lcft{} (and in category theory in general).  There, the Grothendieck ring makes precise the notion of forgetting the indecomposable structure of the modules in the fusion ring, essentially regarding these objects as graded vector spaces (for a precise definition, see \cite[App.~C]{GabFro08}).  Since this is exactly what the characters do, we see that the spirit of the two definitions is the same, and so it is reasonable to call the ring of characters a Grothendieck ring (despite the absence of indecomposable structure in the fusion ring).}.  Assigning modules their characters therefore defines a projection (more precisely, an onto ring homomorphism) from the fusion ring onto the Grothendieck ring.

This has a peculiar effect when considering modular invariance.  Specifically, one expects from rational theories that pairing each module with itself under the holomorphic and antiholomorphic $\func{\affine{\alg{sl}}}{2}$-actions leads to a modular invariant partition function.  But in our case, the coincidence of characters means that there are infinitely many modules all contributing the same amount to the partition function, which therefore diverges.  One can of course regularise this divergence by only allowing the linearly independent characters to contribute, effectively dividing the modular invariant by the infinite multiplicity of each independent character, and in this way one recovers the modular invariant of Kac and Wakimoto \cite{KacMod88b} (we postpone a proper discussion of modularity until \secref{secMod}).  This is indeed invariant under the usual action of $\func{\group{SL}}{2 ; \ZZ}$, but we should be uneasy about its status as a physical partition function.  It does not, strictly speaking, refer to a complete set of modules of the theory.  In particular, there is no set of modules corresponding to this partition function which is closed under fusion.

In essence however, what this does is determine a modular invariant partition function in the Grothendieck ring of characters.  This is no different to what one does in rational theories, and evidence is steadily mounting that this is what one should do in logarithmic theories as well.  However, it is clear that determining a modular invariant in this way does not answer the fundamental question of how the holomorphic and antiholomorphic sectors of the theory are glued together.  For this reason, we advise caution in treating such modular invariants as physical.  Applications require a justification of why such a partition function is appropriate.

We briefly compare this conclusion with that of \cite{LesSU202}.  Their proposal for making sense of Kac and Wakimoto's invariant is to regard the character of $\tfunc{\gamma^{\ell}}{\AffIrrMod{\lambda}}$ as only being defined on the annulus (\ref{eqnAnnulus}).  Summing to get a partition function is therefore viewed as summing over the different annuli in order to have a finite meromorphic partition function on the $z$-plane (with $\abs{q} < 1$).  Presumably this means each character should take value zero outside its given annulus, in defiance of analytic continuation.  Evidence for this proposal is quoted in the claim that a particular modular transformation maps the annuli into one another.  This claim is not true.  Even if it were, the other transformations do not preserve this annulus structure, hence one is \emph{forced} to analytically continue the characters into the rest of the $z$-plane.

We agree that it would be better to extend the definition of partition function so that every module contributes, but the interpretation of \cite{LesSU202} does not achieve this goal.  What is needed in our opinion is an additional quantum number to distinguish representations with the same character.  It is not clear however that such a quantum number need exist.  It seems plausible that modular invariants for fractional level models can only be defined at the level of Grothendieck rings.

\section{Extended Algebras and Ghosts} \label{secGhost}

Note that the fusion rules of $\func{\affine{\alg{sl}}}{2}_{-1/2}$ (\eqnref{eqnFR}) show that every module in the spectrum is a simple current --- these are distinguished in general by the property that fusing them with any irreducible module gives a single irreducible factor \cite{SchSim90}.  This is clear for the modules $\tfunc{\gamma^{\ell}}{\AffIrrMod{0}}$ as they are automorphic images of the vacuum module.  For the other modules, this follows because $\AffIrrMod{1}$ happens to be a simple current.  This is somewhat mysterious as $\AffIrrMod{1}$ is not related to the vacuum by any automorphism of $\func{\affine{\alg{sl}}}{2}$.  Nevertheless, this is the only simple current which has finite order, and its order is $2$ (\eqnref{eqnFR1x1}).

It is therefore interesting to study the \emph{extended algebra} defined by this simple current, more precisely, by the zero-grade fields of the module $\AffIrrMod{1}$.  Their conformal dimension is $\tfrac{1}{2}$ which suggests some sort of fermionic behaviour.  Correctly determining an extended algebra can be a somewhat subtle business, and we shall proceed carefully in an elementary fashion.  The final answer may not be particularly surprising, but there are several pitfalls to avoid during the derivation which we would like to draw attention to.

Let us begin by introducing some convenient notation for the zero-grade fields of the simple current.  We will denote the field corresponding to the \hws{} $\ket{1} \in \AffIrrMod{1}$ by $\phi$ and that corresponding to the descendant $F_0 \ket{1}$ by $\psi$.  We recall \cite{RidSU206} that in general such zero-grade fields are mutually bosonic with respect to the $\func{\affine{\alg{sl}}}{2}$ current field $H$, but mutually fermionic with respect to $E$ and $F$.  This is true even for admissible levels, and the proof is easy in this special case.  Suppose therefore that
\begin{equation} \label{eqnMutLoc}
\func{J}{z} \func{\phi}{w} = \mu_{J,\phi} \func{\phi}{w} \func{J}{z}, \qquad \text{($J = E, H, F$),}
\end{equation}
for some $\mu_{J,\phi} \in \CC$.  Then, we can write
\begin{equation}
\func{J}{z} \func{J'}{x} \func{\phi}{w} = \mu_{J',\phi} \func{J}{z} \func{\phi}{w} \func{J'}{x} = \mu_{J,\phi} \mu_{J',\phi} \func{\phi}{w} \func{J}{z} \func{J'}{x}.
\end{equation}
Alternatively, we can make use of the \ope{} to get
\begin{equation}
\func{J}{z} \func{J'}{x} \func{\phi}{w} = \brac{\frac{\killing{J}{J'} K}{\brac{z-x}^2} + \frac{\func{\comm{J}{J'}}{x}}{z-x} + \ldots} \func{\phi}{w} = \mu_{\sqbrac{J,J'},\phi} \func{\phi}{w} \func{J}{z} \func{J'}{x},
\end{equation}
if $\comm{J}{J'} \neq 0$.  Thus we have
\begin{equation}
\mu_{J,\phi} \mu_{J',\phi} = \mu_{\sqbrac{J,J'},\phi} \qquad \text{if $\comm{J}{J'} \neq 0$.}
\end{equation}
In addition, $K$ remains central in the extended algebra so we can also conclude that
\begin{equation}
\mu_{J,\phi} \mu_{J',\phi} = 1 \qquad \text{if $\killing{J}{J'} \neq 0$.}
\end{equation}

These constraints (which are equivalent to the generalised Jacobi identity at the level of modes) fix $\mu_{H , \phi} = 1$ and $\mu_{E , \phi} \mu_{F , \phi} = 1$.  $H$ and $\phi$ are therefore mutually bosonic, but the case of $E$ or $F$ and $\phi$ is not decided.  The corresponding conclusion for the zero-grade descendant field $\psi$ is identical.  To settle the remaining ambiguity, we need to extend the adjoint (\ref{eqnDefAdj}) to the simple current fields.  Needless to say, the adjoint must define an (antilinear) antiautomorphism on the extended algebra, and it is this requirement that we shall exploit.

Since $\AffIrrMod{1}$ is self-conjugate, the extended adjoint must take the form
\begin{equation}
\phi_n^{\dag} = \eps \psi_{-n} \qquad \Rightarrow \qquad \psi_n^{\dag} = \overline{\eps}^{-1} \phi_{-n},
\end{equation}
where the bar denotes complex conjugation.  We now translate the primary field \opes{} into modes using \eqnref{eqnMutLoc}.  For example, we have
\begin{subequations}
\begin{align}
\func{F}{z} \func{\phi}{w} &= \frac{\func{\psi}{w}}{z-w} + \ldots & &\Rightarrow & F_m \phi_n - \mu_{F , \phi} \phi_n F_m &= \psi_{m+n} \label{eqnI} \\
\text{and} \qquad  \func{E}{z} \func{\psi}{w} &= \frac{\func{\phi}{w}}{z-w} + \ldots & &\Rightarrow & E_m \psi_n - \mu_{E , \psi} \psi_n E_m &= \phi_{m+n}. \label{eqnII} 
\end{align}
\end{subequations}
Taking the adjoint of \eqnref{eqnI} and using $\mu_{E , \phi} \mu_{F , \phi} = 1$, we find that
\begin{equation}
- \abs{\eps}^2 \overline{\mu}_{F , \phi} \brac{E_m \psi_n - \overline{\mu}_{E , \phi} \psi_n E_m} = \phi_{m+n}.
\end{equation}
Comparing with \eqnref{eqnII}, we finally conclude that $\overline{\mu}_{F , \phi} = - \abs{\eps}^{-2}$ and $\overline{\mu}_{E , \phi} = \mu_{E , \psi}$, hence that $\mu_{E , \phi} = \mu_{E , \psi}$ and $\mu_{F , \phi} = \mu_{F , \psi}$ are real and negative.  There are no further constraints to be found, so we are free to choose the most symmetric consistent solution:
\begin{equation}
\eps = 1 \qquad \text{hence} \qquad \mu_{E , \phi} = \mu_{F , \phi} = \mu_{E , \psi} = \mu_{F , \psi} = -1.
\end{equation}
It follows that $E$ and $F$ are both mutually fermionic with respect to $\phi$ and $\psi$, as claimed.

We can now turn to the \opes{} of the simple current fields.  From \eqnref{eqnFR1x1} and conservation of $\func{\alg{sl}}{2}$-weight, we know that these must take the form
\begin{subequations}
\begin{align}
\func{\phi}{z} \func{\phi}{w} &= \alpha \func{E}{w} + \ldots & \func{\psi}{z} \func{\psi}{w} &= \gamma \func{F}{w} + \ldots \\
\func{\phi}{z} \func{\psi}{w} &= \frac{1}{z-w} + \beta \func{H}{w} + \ldots & \func{\psi}{z} \func{\phi}{w} &= \frac{1}{z-w} + \beta' \func{H}{w} + \ldots, \label{eqnOPEsphipsi}
\end{align}
\end{subequations}
for some constants $\alpha$, $\beta$, $\beta'$ and $\gamma$.  These are easily computed.  For example, the first expansion implies that $\phi_{-1/2} \ket{\phi} = \alpha E_{-1} \ket{0}$.  Comparing
\begin{equation}
\bracket{0}{F_1 \phi_{-1/2}}{\phi} = \bracket{0}{\psi_{1/2} \phi_{-1/2}}{0} = 1 \qquad \text{and} \qquad \bracket{0}{F_1 E_{-1}}{0} = \bracket{0}{K - H_0}{0} = \frac{-1}{2}
\end{equation}
immediately yields $\alpha = k^{-1} = -2$.  Similarly, $\beta = -1$, $\beta' = 1$ and $\gamma = -2$.  Note that we have normalised the zero-grade state $\ket{\phi}$ to have norm $1$.  It follows that $\ket{\psi} = F_0 \ket{\phi}$ also has norm $1$ (these norms are the respective constants in the singular term of the \opes{} (\ref{eqnOPEsphipsi})).

We also need to determine the mutual locality of the simple current fields with one another, and this follows easily from the above \opes{}.  For example, if $\func{\phi}{z} \func{\psi}{w} = \mu \func{\psi}{w} \func{\phi}{z}$, then inserting the \opes{} (\ref{eqnOPEsphipsi}) gives
\begin{equation}
\frac{1}{z-w} - \func{H}{w} + \ldots = \mu \brac{\frac{1}{w-z} + \func{H}{z} + \ldots}.
\end{equation}
Taylor-expanding $\func{H}{z}$ about $w$, we see that $\mu = -1$.  Thus, $\phi$ and $\psi$ are mutually fermionic with respect to each other.  Similarly, we can prove that both simple current fields are mutually bosonic with respect to themselves.

Finally, we have to verify that the \opes{} we have derived are \emph{associative}.  For this we need to consider operator products of three fields.  The associativity when at least one of the fields is an affine current is built into the above derivations, so we only need to check the case where all three fields are simple current fields.  For example, since $E$ and $\phi$ are mutually fermionic, we see that
\begin{equation}
\func{\phi}{z} \func{\phi}{w} \func{\phi}{x} = \sqbrac{-2 \func{E}{w} + \ldots} \func{\phi}{x} = - \func{\phi}{x} \sqbrac{-2 \func{E}{w} + \ldots} = - \func{\phi}{x} \func{\phi}{z} \func{\phi}{w}.
\end{equation}
However, this contradicts the fact that $\phi$ is mutually bosonic with respect to itself.  In fact, further computation shows that \emph{every} combination of three simple current fields exhibits the same contradiction --- there is always a lone factor of $-1$ unaccounted for by the mutual locality.

This problem has been observed before in the algebra defining graded parafermions \cite{JacQua06} and certain minimal model extended algebras \cite{RidMin07}.  The remedy is to introduce an auxiliary operator $\mathcal{S}$ which commutes with the affine generators, leaves the vacuum invariant, but \emph{anticommutes} with the simple current fields.  The defining \opes{} of the extended algebra are thereby modified to be
\begin{subequations} \label{eqnExtAlg}
\begin{align}
\func{\phi}{z} \func{\phi}{w} &= \mathcal{S} \sqbrac{-2 \func{E}{w} + \ldots} & \func{\psi}{z} \func{\psi}{w} &= \mathcal{S} \sqbrac{-2 \func{F}{w} + \ldots} \\
\func{\phi}{z} \func{\psi}{w} &= \mathcal{S} \sqbrac{\frac{1}{z-w} - \func{H}{w} + \ldots} & \func{\psi}{z} \func{\phi}{w} &= \mathcal{S} \sqbrac{\frac{1}{z-w} + \func{H}{w} + \ldots}
\end{align}
\end{subequations}
One can check that introducing such an $\mathcal{S}$ precisely accounts for the factor of $-1$ observed above, restoring associativity.  The mutual localities then give the corresponding mode algebra as
\begin{equation} \label{eqnModeAlg}
\comm{\phi_m}{\phi_n} = 0, \qquad \acomm{\phi_m}{\psi_n} = \delta_{m+n,0} \mathcal{S} \qquad \text{and} \qquad \comm{\psi_m}{\psi_n} = 0.
\end{equation}
It is easy to check that $\mathcal{S}$ acts as the identity on each $\tfunc{\gamma^{\ell}}{\AffIrrMod{0}}$, but as \emph{minus} the identity on each $\tfunc{\gamma^{\ell}}{\AffIrrMod{1}}$.  $\mathcal{S}$ is therefore self-inverse, and by \eqnref{eqnModeAlg}, self-adjoint.

Finally, the singular terms of the \opes{} derived here suggest, together with the conformal dimensions of the simple current fields, that what we have constructed is nothing but a complex fermion, or equivalently, a system of fermionic ghosts.  However, this system has central charge $1$ whereas $\func{\affine{\alg{sl}}}{2}_{-1/2}$ has central charge $-1$.  The correct identification is that our extended algebra realises a system of \emph{bosonic} ghosts, given by
\begin{equation}
\func{\beta}{z} = \func{\phi}{z} \qquad \text{and} \qquad \func{\gamma}{z} = \mathcal{S}^{-1} \func{\psi}{z}.
\end{equation}
The defining \opes{} then become
\begin{subequations} \label{eqnGhost}
\begin{align}
\func{\beta}{z} \func{\beta}{w} &= \mathcal{S} \sqbrac{-2 \func{E}{w} + \ldots} & \func{\gamma}{z} \func{\gamma}{w} &= \mathcal{S}^{-1} \sqbrac{2 \func{F}{w} + \ldots} \\
\func{\beta}{z} \func{\gamma}{w} &= \frac{-1}{z-w} + \func{H}{w} + \ldots & \func{\gamma}{z} \func{\beta}{w} &= \frac{1}{z-w} + \func{H}{w} + \ldots
\end{align}
\end{subequations}
It is easy to check that $\beta$ and $\gamma$ are mutually bosonic with respect to themselves and each other.

By making the redefinitions $\widetilde{E} = \mathcal{S} E$, $\widetilde{H} = H$ and $\widetilde{F} = \mathcal{S}^{-1} F$ (which do not affect the affine algebra structure), we recover the standard $\beta \gamma$ ghost \opes{}.  Whilst this trick allows us to remove any trace of $\mathcal{S}$ from the defining equations, and even makes all the fields mutually bosonic with respect to one another, the ghost adjoint still requires the $\mathcal{S}$ operator:
\begin{equation}
\beta^{\dag} = \mathcal{S} \gamma \qquad \text{and} \qquad \gamma^{\dag} = \beta \mathcal{S}^{-1}.
\end{equation}
As the adjoint is vital for computations, we see therefore that we cannot do without $\mathcal{S}$ completely!

\section{A Simplification} \label{secGhost2}

It is worth emphasising once again the fundamental r\^{o}le played by the adjoint (\ref{eqnDefAdj}) in deriving the extended algebra in the previous section.  This is the adjoint corresponding to the real form $\func{\alg{su}}{2}$ of $\func{\alg{sl}}{2} = \func{\alg{sl}}{2 ; \CC}$.  We could also consider the adjoint corresponding to the real form $\func{\alg{sl}}{2 ; \RR}$:
\begin{equation} \label{eqnDefOtherAdj}
J_n^{\ddag} = -J_{-n}, \qquad K^{\ddag} = K \qquad \text{and} \qquad L_n^{\ddag} = L_{-n} \qquad \text{($J = E,H,F$).}
\end{equation}
When we wish to emphasise that the chiral algebra $\func{\affine{\alg{sl}}}{2}$ comes equipped with one of these adjoints, we will denote it by $\func{\affine{\alg{su}}}{2}$ or $\func{\affine{\alg{sl}}}{2 ; \RR}$, as appropriate.  We stress that these are still complex Lie algebras.  In general, every order-$2$ automorphism of a complex simple Lie algebra $\alg{g}$ induces\footnote{If $\Omega$ is the automorphism, the induced adjoint is given by $x^{\dag} = -\func{\Omega}{x}$, where $x$ is either an element of the Cartan subalgebra or a root vector.  This is then extended antilinearly to the entire complex Lie algebra.} an adjoint on $\alg{g}$ and its untwisted affinisation $\affine{\alg{g}}$.  For $\alg{g} = \func{\alg{sl}}{2 ; \CC}$, the adjoint given in \eqnref{eqnDefAdj} corresponds to the non-trivial Weyl reflection whereas that of \eqnref{eqnDefOtherAdj} corresponds to the trivial automorphism.

We want to repeat the derivation of the extended algebra using the $\func{\alg{sl}}{2 ; \RR}$ adjoint.  The result will be slightly different, but the derivation is significantly simpler.  The point here is that the choice of adjoint makes a real difference to simple current extensions of a chiral algebra.  In the theory we are constructing, we have no physical intuition to support either choice, so it is interesting and valid to consider both possibilities.  However, in concrete applications one generally does have a given adjoint, so it is extremely important to be sure that the extended algebra one derives and works with is the correct one.

To proceed, we have to change our basis to something appropriate for $\func{\alg{sl}}{2 ; \RR}$.  The problem here is that the eigenvectors $E$ and $F$ of $\func{\ad}{H}$ are not raising and lowering operators with respect to the adjoint (\ref{eqnDefOtherAdj}).  Instead, we introduce the linear combinations
\begin{equation}
h = \ii \brac{E - F}, \qquad e = \frac{1}{2} \brac{\ii E + \ii F - H} \qquad \text{and} \qquad f = \frac{1}{2} \brac{\ii E + \ii F + H}
\end{equation}
(of $\func{\alg{sl}}{2 ; \CC}$).  These can be quickly checked to satisfy
\begin{equation}
e^{\ddag} = f, \qquad h^{\ddag} = h, \qquad \comm{h}{e} = 2e \qquad \text{and} \qquad \comm{h}{f} = -2f,
\end{equation}
so we have recovered the formalism of raising and lowering operators.  The subtle but important difference between these operators and those considered in \secref{secAlg} is that
\begin{equation}
\comm{e}{f} = -h.
\end{equation}
This difference is mirrored in the Killing form which is given in this basis by (compare \eqnref{eqnKilling})
\begin{equation}
\killing{h}{h} = 2 \qquad \text{and} \qquad \killing{e}{f} = -1,
\end{equation}
with all other entries vanishing.  The basis $\set{e,h,f}$ can now be affinised in the usual manner to define a new basis $\set{e_n,h_n,f_n,K,L_0}$ of $\func{\affine{\alg{sl}}}{2}$.  We will \emph{not} change the normalisation of the central extension $K$ and derivation $L_0$, as compared with \eqnref{eqnCommRels} (we mention this as many articles implicitly replace $K$ by $-K$ which changes the prefactor of $L_0$ in the Sugawara construction).

Consider now the zero-grade states $\ket{\phi}$ and $\ket{\psi}$ of the simple current module $\AffIrrMod{1}$.  Just as we have had to change the basis of algebra generators to account for the $\func{\alg{sl}}{2 ; \RR}$ adjoint, so we need to change this basis.  The problem now is that $\ket{\phi}$ is not a \hws{} with respect to the triangular decomposition afforded by $e_n$, $h_n$ and $f_n$.  Indeed, it is not even an eigenstate of $h_0$.  A better basis of zero-grade states is given by
\begin{equation}
\ket{\Phi} = \ket{\phi} - \ii \ket{\psi} \qquad \text{and} \qquad \ket{\Psi} = \ket{\phi} + \ii \ket{\psi}.
\end{equation}
One can easily check that these are $h_0$-eigenstates with eigenvalues $1$ and $-1$, respectively, and that $e_0 \ket{\Phi} = f_0 \ket{\Psi} = 0$.  Again, there is a subtle difference in the structure:
\begin{equation}
f_0 \ket{\Phi} = \ket{\Psi} \qquad \text{but} \qquad e_0 \ket{\Psi} = - \ket{\Phi}.
\end{equation}
This is reflected in the norms:  If $\ket{\Phi}$ has norm $1$, then $\ket{\Psi}$ has norm $-1$ (the fundamental representation of $\func{\alg{sl}}{2}$ is not unitarisable with respect to the $\func{\alg{sl}}{2 ; \RR}$ adjoint).

We can now repeat the computations of \secref{secGhost}.  First, it is clear that the \opes{} of $e$, $h$ or $f$ with the simple current fields $\Phi$ or $\Psi$ will lead to the same constraints on the mutual localities, namely
\begin{equation}
\mu_{h,\Phi} = \mu_{h,\Psi} = 1 \qquad \text{and} \qquad \mu_{e,\Phi} \mu_{f,\Phi} = \mu_{e,\Psi} \mu_{f,\Psi} = 1.
\end{equation}
We therefore determine when the adjoint extends to an antiautomorphism of the extended algebra.  Defining $\Phi_n^{\ddag} = \eps \Psi_{-n}$, hence $\Psi_n^{\ddag} = \overline{\eps}^{-1} \Phi_{-n}$, we take the adjoint of the algebra relation
\begin{equation}
f_m \Phi_n - \mu_{f,\Phi} \Phi_n f_m = \Psi_{m+n}
\end{equation}
and compare it to the dual relation
\begin{equation}
e_m \Psi_n - \mu_{e,\Psi} \Psi_n e_m = -\Phi_{m+n}
\end{equation}
(note the minus sign!).  This time we find that $\overline{\mu}_{f , \Phi} = \abs{\eps}^{-2}$ and $\overline{\mu}_{e , \Phi} = \mu_{e , \Psi}$, hence that $\mu_{e , \Phi} = \mu_{e , \Psi}$ and $\mu_{f , \Phi} = \mu_{f , \Psi}$ are real and \emph{positive}.  The simplest solution is therefore
\begin{equation}
\eps = 1 \qquad \text{hence} \qquad \mu_{e , \Phi} = \mu_{f , \Phi} = \mu_{e , \Psi} = \mu_{f , \Psi} = 1.
\end{equation}
The algebra generators are therefore mutually bosonic with respect to the simple current fields in this picture!

This is evidently a more familiar situation than that which we found in \secref{secGhost} with the $\func{\alg{su}}{2}$ adjoint.  Continuing with the extended algebra derivation, we can determine the defining \opes{}:
\begin{subequations} \label{eqnNewOPEs}
\begin{align}
\func{\Phi}{z} \func{\Phi}{w} &= 2 \func{e}{w} + \ldots & \func{\Psi}{z} \func{\Psi}{w} &= 2 \func{f}{w} + \ldots \\
\func{\Phi}{z} \func{\Psi}{w} &= \frac{-1}{z-w} + \func{h}{w} + \ldots & \func{\Psi}{z} \func{\Phi}{w} &= \frac{1}{z-w} + \func{h}{w} + \ldots
\end{align}
\end{subequations}
(note that the constants appearing in the singular term of these \opes{} correspond to the respective norms of $\ket{\Psi}$ and $\ket{\Phi}$).  It follows immediately from these expansions that the simple current fields are mutually bosonic with respect to themselves and each other, and it is simple to show that these expansions determine an associative operator product algebra (without any additional $\mathcal{S}$-type operators).  The correspondence with the ghost fields is therefore as natural as it could be:
\begin{equation} \label{eqnNewGhosts}
\func{\beta}{z} = \func{\Phi}{z} \qquad \text{and} \qquad \func{\gamma}{z} = \func{\Psi}{z}.
\end{equation}
Moreover, the adjoint on the ghost fields is just $\beta^{\ddag} = \gamma$ and $\gamma^{\ddag} = \beta$.

It is appropriate now to discuss the reverse procedure, obtaining the $\func{\affine{\alg{sl}}}{2}_{-1/2}$ symmetry from studying the $\beta \gamma$ ghost system, for this was how these theories were first related (see \cite{GurRel98} for example).  From \eqnDref{eqnNewOPEs}{eqnNewGhosts}, we see that the composite fields
\begin{equation}
e = \frac{1}{2} \normord{\beta \beta}, \qquad h = \normord{\beta \gamma} \qquad \text{and} \qquad f = \frac{1}{2} \normord{\gamma \gamma}
\end{equation}
together reconstitute the $\func{\affine{\alg{sl}}}{2}$ generators.  Moreover, explicit calculation confirms that these are the $\func{\alg{sl}}{2 ; \RR}$-type generators of this section.  Furthermore, the ghost adjoint $\beta^{\ddag} = \gamma$ now implies the adjoint (\ref{eqnDefOtherAdj}).

To summarise, the last two sections prove that the $\beta \gamma$ ghost system is naturally a simple current extension of $\func{\affine{\alg{sl}}}{2 ; \RR}_{-1/2}$ (we remind the reader again that we do not negate the level here).  In order to realise these ghosts as an extension of $\func{\affine{\alg{su}}}{2}_{-1/2}$, it is \emph{necessary} to augment the ghost algebra by the operator $\mathcal{S}$.  It is not hard to find examples in the literature where this subtlety has been overlooked, so we want to emphasise the precise results derived here.  Ignoring this leads to contradictions in the algebra when delving deeper into the module structure \cite{RidSU206}.

\section{Extended Algebra Representation Theory} \label{secGhostReps}

We now turn to a discussion of the representations of our extended algebra (\ref{eqnNewOPEs}), or equivalently, of the ghost system (\ref{eqnGhost}) (we work with $\func{\affine{\alg{sl}}}{2 ; \RR}$ for simplicity).  The corresponding algebra relations are
\begin{equation} \label{eqnExtAlgComm}
\comm{\Phi_m}{\Phi_n} = 0, \qquad \comm{\Psi_m}{\Phi_n} = \delta_{m+n,0} \qquad \text{and} \qquad \comm{\Psi_m}{\Psi_n} = 0,
\end{equation}
where we write $\func{\Phi}{z} = \sum_n \Phi_n z^{-n-1/2}$ and $\func{\Psi}{z} = \sum_n \Psi_n z^{-n-1/2}$ as usual.  Since
\begin{equation} \label{eqnFRSC}
\AffIrrMod{1} \fuse \func{\gamma^{\ell}}{\AffIrrMod{0}} = \func{\gamma^{\ell}}{\AffIrrMod{1}} \qquad \text{and} \qquad \AffIrrMod{1} \fuse \func{\gamma^{\ell}}{\AffIrrMod{1}} = \func{\gamma^{\ell}}{\AffIrrMod{0}},
\end{equation}
we see that each (irreducible) extended module will be labelled by a single integer $\ell$, and be composed of two $\func{\affine{\alg{sl}}}{2}_{-1/2}$-modules, $\tfunc{\gamma^{\ell}}{\AffIrrMod{0}}$ and $\tfunc{\gamma^{\ell}}{\AffIrrMod{1}}$.  We denote this extended module by $\ExtIrrMod{\ell}$ ($\ell \in \ZZ$).

To understand the structure of these extended modules, we must first determine their monodromy charges $\theta_{\ell}$.  These determine whether the extended algebra modes $\Phi_n$ and $\Psi_n$ have indices which are integers, half-integers, or something else entirely.  The monodromy charge may be defined \cite{SchSim90,RidSU206} in terms of the powers of $z-w$ appearing in the \opes{} of the simple current fields $\func{\Phi}{z}$ and $\func{\Psi}{z}$ with a field $\func{\xi^{\brac{\ell}}}{z}$ associated to a state of the extended module $\ExtIrrMod{\ell}$.  The simple current property means that the powers of $z-w$ which appear only differ by an integer, and their common value \emph{modulo} $\ZZ$ defines the monodromy charge (strictly speaking, this is the negative of the monodromy charge).  This is also obviously independent of the choice of $\xi^{\brac{\ell}}$ and simple current field.

From \eqnref{eqnTransAuts}, we can compute (as in \secref{secFusion}) that $\tfunc{\gamma^{\ell}}{\ket{0}}$ has conformal dimension $-\ell^2 / 8$ whereas that of $\tfunc{\gamma^{\ell}}{\ket{1}}$ is $-\brac{\ell^2 - 4 \ell - 4} / 8$.  The fusion rules (\ref{eqnFRSC}) then imply that the monodromy charges of $\tfunc{\gamma^{\ell}}{\AffIrrMod{0}}$ and $\tfunc{\gamma^{\ell}}{\AffIrrMod{1}}$ are
\begin{equation}
\tfrac{1}{2} - \tfrac{1}{8} \ell^2 + \tfrac{1}{8} \brac{\ell^2 - 4 \ell - 4} = \tfrac{-1}{2} \ell \qquad \text{and} \qquad \tfrac{1}{2} - \tfrac{1}{8} \brac{\ell^2 - 4 \ell - 4} + \tfrac{1}{8} \ell^2 = \tfrac{1}{2} \ell + 1,
\end{equation}
respectively.  The monodromy charge of the extended module $\ExtIrrMod{\ell}$ is therefore well-defined (as claimed) and is simply
\begin{equation}
\theta_{\ell} = \frac{\ell}{2} \pmod{\ZZ}.
\end{equation}
It now follows that when $\func{\Phi}{z}$ and $\func{\Psi}{z}$ act upon a state $\ket{\xi^{\brac{\ell}}} \in \ExtIrrMod{\ell}$ (with monodromy charge $\theta_{\ell}$), they must be expanded in the forms
\begin{equation}
\func{\Phi}{z} \ket{\xi^{\brac{\ell}}} = \sum_{n \in \ZZ + \theta_{\ell} - 1/2} \Phi_n z^{-n - 1/2} \ket{\xi^{\brac{\ell}}} \qquad \text{and} \qquad \func{\Psi}{z} \ket{\xi^{\brac{\ell}}} = \sum_{n \in \ZZ + \theta_{\ell} - 1/2} \Psi_n z^{-n - 1/2} \ket{\xi^{\brac{\ell}}}.
\end{equation}
In other words, the modes $\Phi_n$ and $\Psi_n$ with $n \in \ZZ + \tfrac{1}{2}$ act on the extended algebra modules $\ExtIrrMod{\ell}$ with $\ell \in 2 \ZZ$, and the modes $\Phi_n$ and $\Psi_n$ with $n \in \ZZ$ act on the extended algebra modules $\ExtIrrMod{\ell}$ with $\ell \in 2 \ZZ + 1$.

For example, the extended vacuum module $\ExtIrrMod{0} \sim \AffIrrMod{0} \oplus \AffIrrMod{1}$ has monodromy charge $0$, so $\Phi_n$ and $\Psi_n$ act upon it with half-integer indices.  In particular, $\Phi_{-1/2}$ and $\Psi_{-1/2}$ act on the vacuum to create the zero-grade states $\ket{\Phi}$ and $\ket{\Psi}$ of the simple current module $\AffIrrMod{1}$ (these are not the same as $\ket{1} = \ket{\phi}$ and $F_0 \ket{1} = \ket{\psi}$ as we changed basis in \secref{secGhost2}).  We recall from \secref{secFusion} that every module in our $\func{\affine{\alg{sl}}}{2}_{-1/2}$ theory could be regarded as the image under a spectral flow automorphism of either $\AffIrrMod{0}$ or $\AffIrrMod{1}$.  It is reasonable therefore to expect that the same conclusion will hold for the extended algebra modules.

This is indeed the case.  The spectral flow\footnote{We denote this spectral flow by $\widetilde{\gamma}$ because this automorphism is not the same as the spectral flow automorphism $\gamma$ which was introduced in \eqnref{eqnSF}.  Whereas the latter denotes a spectral flow naturally defined on the $\func{\affine{\alg{su}}}{2}$ basis, the flow $\widetilde{\gamma}$ is naturally defined on the $\func{\affine{\alg{sl}}}{2 ; \RR}$ basis.  We can see that these are different by determining the action of $\widetilde{\gamma}$ on the $\func{\affine{\alg{su}}}{2}$ basis:
\begin{align*}
\func{\widetilde{\gamma}}{E_n} &= \frac{1}{4} \brac{E_{n-1} + 2 E_n + E_{n+1} + \ii H_{n-1} + 2 \ii \delta_{n,0} K - \ii H_{n+1} + F_{n-1} - 2 F_n + F_{n+1}} \\
\func{\widetilde{\gamma}}{H_n} &= \frac{1}{2} \brac{-\ii E_{n-1} + \ii E_{n+1} + H_{n-1} + H_{n+1} - \ii F_{n-1} + \ii F_{n+1}} \\
\func{\widetilde{\gamma}}{F_n} &= \frac{1}{4} \brac{E_{n-1} - 2 E_n + E_{n+1} + \ii H_{n-1} - 2 \ii \delta_{n,0} K - \ii H_{n+1} + F_{n-1} + 2 F_n + F_{n+1}}.
\end{align*}
While this can be checked to indeed provide a non-trivial automorphism (of $\func{\affine{\alg{sl}}}{2 ; \CC}$), it is not clear whether it is of any use in further analysing our theory.  Note that it preserves the adjoint (\ref{eqnDefOtherAdj}) but not (\ref{eqnDefAdj}).}
\begin{subequations} \label{eqnSFsl2R}
\begin{gather}
\func{\widetilde{\gamma}}{e_n} = e_{n-1} \qquad \func{\widetilde{\gamma}}{h_n} = h_n - \delta_{n,0} K \qquad \func{\widetilde{\gamma}}{f_n} = f_{n+1} \\
\func{\widetilde{\gamma}}{K} = K \qquad \func{\widetilde{\gamma}}{L_0} = L_0 - \frac{1}{2} h_0 + \frac{1}{4} K
\end{gather}
\end{subequations}
may be derived from the following extended algebra automorphism (which we also denote by $\widetilde{\gamma}$)
\begin{equation} \label{eqnExtSF}
\func{\widetilde{\gamma}}{\Phi_n} = \Phi_{n - 1/2} \qquad \func{\widetilde{\gamma}}{\Psi_n} = \Psi_{n + 1/2}.
\end{equation}
Glancing at \eqnref{eqnExtAlgComm}, this is obviously an extended algebra automorphism, and the change of mode indices from integer to half-integer and vice-versa precisely accounts for the fact that the monodromy charge changes in this way when applying the $\func{\affine{\alg{sl}}}{2}$ spectral flow.

To show that \eqnref{eqnExtSF} implies \eqnref{eqnSFsl2R}, we derive certain generalised commutation relations relating extended algebra modes and affine modes.  These are obtained by evaluating
\begin{equation}
\oint_0 \oint_w \func{\Phi}{z} \func{\Phi}{w} z^{m+1/2} w^{n-1/2} \brac{z-w}^{-1} \frac{\dd z}{2 \pi \ii} \frac{\dd w}{2 \pi \ii}
\end{equation}
in two different ways (and by replacing one or both of the fields $\Phi$ by $\Psi$).  We can expand the operator product directly, using \eqnref{eqnNewOPEs}, or we can break the $z$-contour around $w$ into the difference of two contours around the origin, one with $\abs{z} > \abs{w}$ and the other with $\abs{z} < \abs{w}$.  The results of this procedure are the following generalised commutation relations:
\begin{subequations} \label{eqnGCRs}
\begin{align}
\sum_{j=0}^{\infty} \sqbrac{\Phi_{m-j} \Phi_{n+j} + \Phi_{n-j-1} \Phi_{m+j+1}} &= 2 e_{m+n}, \label{eqnGCRe} \\
\sum_{j=0}^{\infty} \sqbrac{\Psi_{m-j} \Phi_{n+j} + \Phi_{n-j-1} \Psi_{m+j+1}} &= h_{m+n} + \brac{m + \frac{1}{2}} \delta_{m+n,0}, \label{eqnGCRh} \\
\sum_{j=0}^{\infty} \sqbrac{\Psi_{m-j} \Psi_{n+j} + \Psi_{n-j-1} \Psi_{m+j+1}} &= 2 f_{m+n}. \label{eqnGCRf}
\end{align}
\end{subequations}
Regarding these as defining relations for the affine modes\footnote{We mention that this is the correct way of defining these modes given the \opes{} (\ref{eqnNewOPEs}), despite the fact that $m$ can be chosen arbitrarily (up to monodromy charge considerations).  Na\"{\i}vely defining the affine modes as the obvious normally-ordered products of the extended algebra modes gives equivalent results, \emph{except} for $h_0$ when $m \in \ZZ$.  Then the na\"{\i}ve result is incorrect, and must be adjusted by the appropriate multiple of the identity.  This correction phenomenon should be familiar from the computation of the Virasoro zero-mode in the Ramond sector of the free fermion.}, it is easy to check that applying \eqnref{eqnExtSF} recovers the affine spectral flow (with the implicit replacement of $K$ by $k = \tfrac{-1}{2}$).

Now consider the singular vectors of the extended algebra module $\ExtIrrMod{0}$, or rather of the corresponding Verma module $\ExtVerMod{0}$.  Since one expects this module to be composed of the $\func{\affine{\alg{sl}}}{2}_{-1/2}$ Verma modules $\AffVerMod{0}$ and $\AffVerMod{1}$, there are four non-trivial singular vector combinations to consider:
\begin{subequations}
\begin{gather}
f_0 \ket{0} \in \AffVerMod{0}, \qquad f_0 \ket{\Psi} \in \AffVerMod{1}, \\
\brac{156 e_{-3} e_{-1} - 71 e_{-2}^2 + 44 e_{-2} h_{-1} e_{-1} - 52 h_{-2} e_{-1}^2 + 16 f_{-1} e_{-1}^3 - 4 h_{-1}^2 e_{-1}^2} \ket{0} \in \AffVerMod{0} \label{eqnSV0} \\
\text{and} \qquad \brac{7 e_{-2} - 2 h_{-1} e_{-1}} \ket{\Phi} + 4 e_{-1}^2 \ket{\Psi} \in \AffVerMod{1}. \label{eqnSV1}
\end{gather}
\end{subequations}
Note the slight sign change in (\ref{eqnSV0}) as compared to (\ref{eqnVacSV}) due to our change of basis.  Note also that (\ref{eqnSV1}) has the correct dimension and $\func{\alg{sl}}{2}$-weight as given in \eqnref{eqnV1SVs}.

But applying \eqnref{eqnGCRf} with $m = \tfrac{-1}{2}$ to $f_0 \ket{0}$ gives
\begin{equation}
f_0 \ket{0} = \sum_{j=0}^{\infty} \Psi_{-j - 1/2} \Psi_{j + 1/2} \ket{0} = 0,
\end{equation}
since there are no states in $\ExtVerMod{0}$ with conformal dimension less than $0$.  Similarly,
\begin{equation}
f_0 \ket{\Psi} = f_0 \Psi_{-1/2} \ket{0} = \Psi_{-1/2} f_0 \ket{0} = 0.
\end{equation}
We therefore see that $f_0 \ket{0}$ and $f_0 \ket{\Psi}$ are not (non-trivial) singular vectors in $\ExtVerMod{0}$, rather they vanish \emph{identically}.  It is somewhat more surprising that the same is true for the vectors (\ref{eqnSV0}) and (\ref{eqnSV1}).  We will detail this computation for the latter vector leaving the former as a simple if tedious exercise.

Consider therefore the first term of (\ref{eqnSV1}), $e_{-2} \ket{\Phi} = e_{-2} \Phi_{-1/2} \ket{0}$.  Commuting the affine mode to the right and using \eqnref{eqnGCRe} with $m = \tfrac{-1}{2}$ gives
\begin{equation}
e_{-2} \ket{\Phi} = \frac{1}{2} \Phi_{-1/2} \brac{\Phi_{-1/2} \Phi_{-3/2} + \Phi_{-3/2} \Phi_{-1/2}} \ket{0} = \Phi_{-3/2} \Phi_{-1/2}^2 \ket{0}.
\end{equation}
Repeating this process with $e_{-1} \ket{\Phi}$ and then $h_{-1} e_{-1} \ket{\Phi}$ (using \eqnref{eqnGCRh}) yields
\begin{equation}
h_{-1} e_{-1} \ket{\Phi} = \brac{\frac{3}{2} \Phi_{-3/2} \Phi_{-1/2}^2  + \frac{1}{2} \Psi_{-1/2} \Phi_{-1/2}^4} \ket{0}.
\end{equation}
Finally, recalling that $\comm{e_m}{\Psi_n} = -\Phi_{m+n}$, we derive that
\begin{equation}
e_{-1}^2 \ket{\Psi} = \brac{\frac{1}{4} \Psi_{-1/2} \Phi_{-1/2}^4 - \Phi_{-3/2} \Phi_{-1/2}^2} \ket{0}.
\end{equation}
We therefore see that all the terms of (\ref{eqnSV1}) explicitly cancel, hence that this singular vector also vanishes identically in $\ExtVerMod{0}$.

It follows from the identical vanishing of these singular vectors that the extended algebra Verma module $\ExtVerMod{0}$ is irreducible and is therefore composed of irreducible $\func{\affine{\alg{sl}}}{2}_{-1/2}$-modules:  $\ExtVerMod{0} = \ExtIrrMod{0} \sim \AffIrrMod{0} \oplus \AffIrrMod{1}$.  Because the $\func{\affine{\alg{sl}}}{2}$ spectral flow lifts to a spectral flow automorphism (\ref{eqnExtSF}) on the extended algebra, we may immediately deduce that the other extended algebra modules $\ExtVerMod{\ell} = \tfunc{\widetilde{\gamma}^{\ell}}{\ExtVerMod{0}}$ ($\ell \in \ZZ$), which will not be Verma modules in general, are likewise irreducible.  We mention that the irreducibility of extended algebra Verma modules is generic for (finite) simple current extensions \cite{RidSU206,RidMin07}, although the extended algebra will usually have to be defined by generalised commutation relations.

The extended algebra characters are therefore easily deduced from the obvious Verma module (Poincar\'{e}-Birkhoff-Witt) bases.  Indeed, the character of the extended vacuum module is just
\begin{equation} \label{eqnExtChar0}
\ch{\ExtIrrMod{0}}{z ; q} = \prod_{i=1}^{\infty} \frac{1}{\brac{1 - z^{-1} q^{i-1/2}} \brac{1 - z q^{i-1/2}}} = \sum_{n \in \ZZ / 2} \sum_{m = \abs{n}}^{\infty} \frac{q^m}{\qfact{q}{m-n} \qfact{q}{m+n}} z^{2n},
\end{equation}
where $\qfact{q}{m} = \prod_{i=1}^m \brac{1 - q^i}$ as usual, and we have used the well-known partition identity \cite[Eq.~2.2.5]{AndThe76}
\begin{equation}
\prod_{i=1}^{\infty} \frac{1}{1 - z q^i} = \sum_{j=0}^{\infty} \frac{q^j}{\qfact{q}{j}} z^j.
\end{equation}
This is an example of a so-called \emph{fermionic character formula} --- upon expanding the $\qfact{q}{m}$ factors in the denominator, we find that all the contributions to the sums come with positive signs.  Splitting the sum over $n$ into $n \in \ZZ$ and $n \in \ZZ + \tfrac{1}{2}$ gives fermionic character formulae for the affine modules $\AffIrrMod{0}$ and $\AffIrrMod{1}$, respectively.  This is to be contrasted with the \emph{bosonic} character formulae given for these modules in \eqnDref{eqnCh0}{eqnCh1} which are not manifestly positive in this sense.  The difference is that before we had to subtract and add contributions corresponding to the braiding pattern of the $\func{\affine{\alg{sl}}}{2}_{-1/2}$ singular vectors (\figref{figVacVerMod}).  In the extended algebra picture, these singular vectors all vanish identically, leading to far nicer, manifestly positive character formulae.

Applying the spectral flow one more time, we get expressions for the characters of the extended modules $\ExtIrrMod{\ell}$:
\begin{equation}
\ch{\ExtIrrMod{\ell}}{z ; q} = \frac{z^{-\ell / 2} q^{-\ell^2 / 8}}{\displaystyle \prod_{i=1}^{\infty} \brac{1 - z^{-1} q^{i - \brac{\ell + 1}/2}} \brac{1 - z q^{i + \brac{\ell - 1}/2}}} = z^{-\ell / 2} q^{-\ell^2 / 8} \sum_{n \in \ZZ / 2} \sum_{m = \abs{n}}^{\infty} \frac{q^{m + \ell n}}{\qfact{q}{m-n} \qfact{q}{m+n}} z^{2n}.
\end{equation}
The product forms tell us directly (compare \secref{secChar}) that these characters have simple poles when $z^2 = q^i$ for all $i \in 2 \ZZ - 1 - \ell$.  The fermionic sum form is even nicer.  It gives the decomposition of the character into so-called string functions of constant $\func{\alg{sl}}{2}$-weight.  Unlike the $q$-expansions of \secref{secChar}, these string functions have $q$-expansions which always give the multiplicities of the weights of the modules correctly.  For example, when $\ell = 2$ the terms with $\func{\alg{sl}}{2}$-weight $2n$ have $q$-expansion $q^{\abs{n} + 2n} + \ldots$, so the lowest power of $q$ is $3n > 0$ when $n$ is positive, but is $n < 0$ when $n$ is negative (compare with the depictions of the affine modules in \figref{figSpecFlow}).  Again, restricting the sum to $n$ integer or half-integer recovers fermionic character formulae for the constituent $\func{\affine{\alg{sl}}}{2}_{-1/2}$-modules.

\section{Modular Invariance} \label{secMod}

Finally, we consider the modular properties of the $\func{\affine{\alg{sl}}}{2}_{-1/2}$-characters.  Whereas the bosonic character formulae (\ref{eqnCharSF0}) and (\ref{eqnCharSF1}) for the affine modules were naturally expressed in terms of classical theta functions, the characters of the extended algebra may be expressed in terms of ordinary Jacobi theta functions (our conventions for these are summarised in \appref{appTheta}).  Before giving these expressions, it is convenient to redefine the characters (in the standard manner) by
\begin{equation}
\nch{\IndMod{}}{y ; z ; q} = \tr_{\IndMod{}} y^K z^{H_0} q^{L_0 - C/24}.
\end{equation}
Since $C$ and $K$ are central, the only effect of this redefinition is to multiply the characters by the factors $q^{-c/24} = q^{1/24}$ and $y^k = y^{-1/2}$.  This may seem trivial, especially the inclusion of the new variable $y$, but is in fact essential for constructing representations of the modular group $\func{\group{SL}}{2 ; \ZZ}$ \cite{KacInf84}.

To begin, let us compare \eqnDref{eqnThIdPF}{eqnDefeta} with the product form of the character formula (\ref{eqnExtChar0}).  We find that
\begin{equation}
\nch{\ExtIrrMod{0}}{y ; z ; q} = y^{-1/2} \frac{\func{\eta}{q}}{\Jth{4}{z ; q}}.
\end{equation}
As the $\func{\alg{sl}}{2}$-weights of $\AffIrrMod{0}$ are all even whereas those of $\AffIrrMod{1}$ are all odd, we can project onto the affine characters using the known behaviour of the theta functions under $z \rightarrow e^{\ii \pi} z$ (\eqnref{eqnThId-z}):
\begin{subequations}
\begin{align}
\nch{\AffIrrMod{0}}{y ; z ; q} &= \frac{y^{-1/2}}{2} \sqbrac{\frac{\func{\eta}{q}}{\Jth{4}{z ; q}} + \frac{\func{\eta}{q}}{\Jth{3}{z ; q}}} &
\nch{\AffIrrMod{1}}{y ; z ; q} &= \frac{y^{-1/2}}{2} \sqbrac{\frac{\func{\eta}{q}}{\Jth{4}{z ; q}} - \frac{\func{\eta}{q}}{\Jth{3}{z ; q}}}.
\end{align}
Spectral flow and \eqnref{eqnThIdSF} then give
\begin{align}
\nch{\tfunc{\gamma}{\AffIrrMod{0}}}{y ; z ; q} &= \frac{y^{-1/2}}{2} \sqbrac{\frac{-\ii \func{\eta}{q}}{\Jth{1}{z ; q}} + \frac{\func{\eta}{q}}{\Jth{2}{z ; q}}} &
\nch{\tfunc{\gamma}{\AffIrrMod{1}}}{y ; z ; q} &= \frac{y^{-1/2}}{2} \sqbrac{\frac{-\ii \func{\eta}{q}}{\Jth{1}{z ; q}} - \frac{\func{\eta}{q}}{\Jth{2}{z ; q}}}.
\end{align}
\end{subequations}
These are the four linearly independent (admissible) characters of our theory.

It is now clear from \eqnDref{eqnThIdS}{eqnetaS} that the action of the modular transformation $S$ on the ratios $\eta / \vartheta_i$ appearing in the admissible characters will be to recover such a ratio, but multiplied by the factor $\func{\exp}{-\ii \pi \zeta^2 / \tau}$, where $z = \func{\exp}{2 \pi \ii \zeta}$ and $q = \func{\exp}{2 \pi \ii \tau}$.  Cancelling this unwanted factor is the reason why we must include the variable $y$ in the normalised characters.  Specifically, if $y = \func{\exp}{2 \pi \ii t}$, then we can extend the action (\ref{eqnModGenAct}) of the modular group generators as follows:
\begin{equation}
S \colon \brac{t , \zeta , \tau} \longmapsto \brac{t - \zeta^2 / \tau , \zeta / \tau , -1 / \tau} \qquad T \colon \brac{t , \zeta , \tau} \longmapsto \brac{t , \zeta , \tau + 1}.
\end{equation}
One can easily check that $S^4 = \brac{ST}^6 = \id$ as before.  With this extended action, we can now compute (in hopefully obvious notation)
\begin{align}
\nch{\AffIrrMod{0}}{t - \zeta^2 / \tau \mid \zeta / \tau \mid -1 / \tau} &= \frac{e^{-\ii \pi t}}{2} \sqbrac{\frac{\func{\eta}{\tau}}{\Jth{2}{\zeta \mid \tau}} + \frac{\func{\eta}{\tau}}{\Jth{3}{\zeta \mid \tau}}} \notag \\
&= \frac{1}{2} \sqbrac{\widetilde{\chi}_{\tfunc{\gamma}{\AffIrrMod{0}}} - \widetilde{\chi}_{\tfunc{\gamma}{\AffIrrMod{1}}} + \widetilde{\chi}_{\AffIrrMod{0}} - \widetilde{\chi}_{\AffIrrMod{1}}} \brac{t \mid \zeta \mid \tau} \\[1mm]
\nch{\AffIrrMod{0}}{t \mid \zeta \mid \tau + 1} &= e^{\ii \pi / 12} \nch{\AffIrrMod{0}}{t \mid \zeta \mid \tau}.
\end{align}

Repeating these computations for the other admissible characters, we obtain the $S$-matrix and $T$-matrix representing these modular transformations on the vector space spanned by the admissible characters.  With respect to the ordered basis
\begin{equation} \label{eqnAdmChBasis}
\set{\widetilde{\chi}_{\AffIrrMod{0}}, \widetilde{\chi}_{\AffIrrMod{1}}, \widetilde{\chi}_{\tfunc{\gamma}{\AffIrrMod{0}}}, \widetilde{\chi}_{\tfunc{\gamma}{\AffIrrMod{1}}}}
\end{equation}
(which corresponds to the admissible \hwms{}), these matrices are
\begin{equation}
S = \frac{1}{2} 
\begin{pmatrix}
1 & -1 & 1 & -1 \\
-1 & 1 & 1 & -1 \\
1 & 1 & \ii & \ii \\
-1 & -1 & \ii & \ii
\end{pmatrix}
\qquad \text{and} \qquad
T = 
\begin{pmatrix}
e^{\ii \pi / 12} & 0 & 0 & 0 \\
0 & -e^{\ii \pi / 12} & 0 & 0 \\
0 & 0 & e^{-\ii \pi / 6} & 0 \\
0 & 0 & 0 & e^{-\ii \pi / 6}
\end{pmatrix}
.
\end{equation}
Both matrices are symmetric and unitary.  We note that $S^2 \colon \brac{t , \zeta , \tau} \longmapsto \brac{t , -\zeta , \tau}$ represents conjugation, but that
\begin{equation} \label{eqnS^2=C}
S^2 = 
\begin{pmatrix}
1 & 0 & 0 & 0 \\
0 & 1 & 0 & 0 \\
0 & 0 & 0 & -1 \\
0 & 0 & -1 & 0
\end{pmatrix}
.
\end{equation}
This indicates that $\AffIrrMod{0}$ and $\AffIrrMod{1}$ are self-conjugate, as we know, but the appearance of the negative entries in the last two rows deserves comment.  These negative entries may be explained by noting that the conjugates of the \hwms{} $\tfunc{\gamma}{\AffIrrMod{0}}$ and $\tfunc{\gamma}{\AffIrrMod{1}}$ are the non-\hwms{} $\tfunc{\gamma^{-1}}{\AffIrrMod{0}}$ and $\tfunc{\gamma^{-1}}{\AffIrrMod{1}}$ (respectively).  The latter modules do not appear in the list of admissible modules, but their characters satisfy (\secref{secChar})
\begin{equation}
\nch{\tfunc{\gamma^{-1}}{\AffIrrMod{0}}}{y ; z ; q} = -\nch{\tfunc{\gamma}{\AffIrrMod{1}}}{y ; z ; q} \qquad \text{and} \qquad \nch{\tfunc{\gamma^{-1}}{\AffIrrMod{1}}}{y ; z ; q} = -\nch{\tfunc{\gamma}{\AffIrrMod{0}}}{y ; z ; q}.
\end{equation}
This precisely accounts for the negative off-diagonal entries in $S^2$.  Put differently, this shows that $S^2$ represents conjugation on the Grothendieck ring of characters (\secref{secChar}).

The diagonal modular invariant therefore takes the form
\begin{align}
\func{\mathcal{Z}_{\text{diag.}}}{y ; z ; q} &= \tabs{\widetilde{\chi}_{\AffIrrMod{0}}}^2 + \tabs{\widetilde{\chi}_{\AffIrrMod{1}}}^2 + \tabs{\widetilde{\chi}_{\tfunc{\gamma}{\AffIrrMod{0}}}}^2 + \tabs{\widetilde{\chi}_{\tfunc{\gamma}{\AffIrrMod{1}}}}^2 \notag \\
&= \frac{1}{2 \abs{y}} \sqbrac{\frac{\abs{\func{\eta}{q}}^2}{\abs{\Jth{4}{z ; q}}^2} + \frac{\abs{\func{\eta}{q}}^2}{\abs{\Jth{3}{z ; q}}^2} + \frac{\abs{\func{\eta}{q}}^2}{\abs{\Jth{2}{z ; q}}^2} + \frac{\abs{\func{\eta}{q}}^2}{\abs{\Jth{1}{z ; q}}^2}}.
\end{align}
Furthermore, \eqnref{eqnS^2=C} specifies that the charge-conjugate modular invariant takes the form
\begin{align}
\func{\mathcal{Z}_{\text{cc.}}}{y ; z ; q} &= \tabs{\widetilde{\chi}_{\AffIrrMod{0}}}^2 + \tabs{\widetilde{\chi}_{\AffIrrMod{1}}}^2 - \widetilde{\chi}_{\tfunc{\gamma}{\AffIrrMod{0}}} \widetilde{\chi}_{\tfunc{\gamma}{\AffIrrMod{1}}}^* - \widetilde{\chi}_{\tfunc{\gamma}{\AffIrrMod{1}}} \widetilde{\chi}_{\tfunc{\gamma}{\AffIrrMod{0}}}^* \notag \\
&= \tabs{\widetilde{\chi}_{\AffIrrMod{0}}}^2 + \tabs{\widetilde{\chi}_{\AffIrrMod{1}}}^2 + \widetilde{\chi}_{\tfunc{\gamma}{\AffIrrMod{0}}} \widetilde{\chi}_{\tfunc{\gamma^{-1}}{\AffIrrMod{0}}}^* + \widetilde{\chi}_{\tfunc{\gamma}{\AffIrrMod{1}}}
\widetilde{\chi}_{\tfunc{\gamma^{-1}}{\AffIrrMod{1}}}^* \notag \\
&= \frac{1}{2 \abs{y}} \sqbrac{\frac{\abs{\func{\eta}{q}}^2}{\abs{\Jth{4}{z ; q}}^2} + \frac{\abs{\func{\eta}{q}}^2}{\abs{\Jth{3}{z ; q}}^2} + \frac{\abs{\func{\eta}{q}}^2}{\abs{\Jth{2}{z ; q}}^2} - \frac{\abs{\func{\eta}{q}}^2}{\abs{\Jth{1}{z ; q}}^2}},
\end{align}
where the asterisks denote complex conjugation.  We emphasise the negative coefficients appearing with respect to the basis (\ref{eqnAdmChBasis}).  If one neglects these signs (as in \cite{LesSU202}), then the ``invariant'' transforms non-trivially under the modular $S$ transformation.  Indeed, it is not hard to show that every modular invariant must have the form
\begin{equation}
\func{\mathcal{Z}_m}{y ; z ; q} = \func{\mathcal{Z}_{\text{diag.}}}{y ; z ; q} + m \tabs{\widetilde{\chi}_{\tfunc{\gamma}{\AffIrrMod{0}}} + \widetilde{\chi}_{\tfunc{\gamma}{\AffIrrMod{1}}}}^2, \qquad m \in \ZZ.
\end{equation}
(In this classification, $\mathcal{Z}_{\text{diag.}} = \mathcal{Z}_0$ and $\mathcal{Z}_{\text{cc.}} = \mathcal{Z}_{-1}$.)  This reflects the simple observation that
\begin{equation}
\nch{\tfunc{\gamma}{\AffIrrMod{0}}}{y ; z ; q} + \nch{\tfunc{\gamma}{\AffIrrMod{1}}}{y ; z ; q} = \nch{\ExtIrrMod{1}}{y ; z ; q} = -\ii y^{-1/2} \frac{\func{\eta}{q}}{\Jth{1}{z ; q}}
\end{equation}
is itself $\func{\group{SL}}{2 ; \ZZ}$-invariant, up to a factor of $\ii$.

Finally, it is appropriate to discuss the Verlinde formula.  In rational theories, this summarises a remarkable connection between the modular properties of the characters and the fusion ring.  If $\mathcal{N}_{\lambda \mu}^{\hphantom{\lambda \mu} \nu}$ denotes the multiplicity with which $\AffIrrMod{\nu}$ appears in the fusion decomposition of $\AffIrrMod{\lambda}$ and $\AffIrrMod{\mu}$, then the Verlinde formula relates these fusion multiplicities to the modular $S$-matrix via
\begin{equation}
\mathcal{N}_{\lambda \mu}^{\hphantom{\lambda \mu} \nu} = \sum_{\sigma} \frac{S_{\lambda \sigma} S_{\mu \sigma} S_{\nu \sigma}^*}{S_{0 \sigma}}.
\end{equation}
Here the sum runs over all irreducible modules $\AffIrrMod{\sigma}$ in the fusion ring, and the index $0$ refers to the vacuum module $\AffIrrMod{0}$.

In our fractional level theory, we no longer have a bijective correspondence between the modules of the theory and the characters, so it is pointless to expect a direct relation between the fusion ring of our theory and the $S$-matrix.  However, we can compute the ``fusion multiplicities'' obtained from the Verlinde formula by restricting the sum to the linearly independent admissible characters (\ref{eqnAdmChBasis}).  Collecting these multiplicities in fusion matrices, $\brac{N_{\lambda}}_{\mu \nu} = \mathcal{N}_{\lambda \mu}^{\hphantom{\lambda \mu} \nu}$, the results are
\begin{subequations}
\begin{gather}
N_{\AffIrrMod{0}} = 
\begin{pmatrix}
1 & 0 & 0 & 0 \\
0 & 1 & 0 & 0 \\
0 & 0 & 1 & 0 \\
0 & 0 & 0 & 1
\end{pmatrix}
\qquad N_{\AffIrrMod{1}} = 
\begin{pmatrix}
0 & 1 & 0 & 0 \\
1 & 0 & 0 & 0 \\
0 & 0 & 0 & 1 \\
0 & 0 & 1 & 0
\end{pmatrix}
\\
N_{\tfunc{\gamma}{\AffIrrMod{0}}} = 
\begin{pmatrix}
0 & 0 & 1 & 0 \\
0 & 0 & 0 & 1 \\
0 & -1 & 0 & 0 \\
-1 & 0 & 0 & 0
\end{pmatrix}
\qquad N_{\tfunc{\gamma}{\AffIrrMod{1}}} = 
\begin{pmatrix}
0 & 0 & 0 & 1 \\
0 & 0 & 1 & 0 \\
-1 & 0 & 0 & 0 \\
0 & -1 & 0 & 0
\end{pmatrix}
.
\end{gather}
\end{subequations}
Whilst the negative ``fusion'' multiplicities might seem alarming at first sight, it is easy to check that these are precisely the structure constants of the Grothendieck ring of characters.  For example, the Verlinde formula gives 
\begin{equation}
\mathcal{N}_{\tfunc{\gamma}{\AffIrrMod{1}} \tfunc{\gamma}{\AffIrrMod{0}}}^{\hphantom{\tfunc{\gamma}{\AffIrrMod{1}} \tfunc{\gamma}{\AffIrrMod{0}}} \AffIrrMod{0}} = -1,
\end{equation}
which reflects the Grothendieck fusion rule
\begin{equation}
\widetilde{\chi}_{\tfunc{\gamma}{\AffIrrMod{1}}} \fuse \widetilde{\chi}_{\tfunc{\gamma}{\AffIrrMod{0}}} = -\widetilde{\chi}_{\AffIrrMod{0}}.
\end{equation}
This is of course the projection of the fusion rule
\begin{equation}
\tfunc{\gamma}{\AffIrrMod{1}} \fuse \tfunc{\gamma}{\AffIrrMod{0}} = \tfunc{\gamma^2}{\AffIrrMod{1}}
\end{equation}
onto the characters, by \eqnref{eqnLinIndChars}.  There is no mystery here --- the modular $S$-matrix only sees the Grothendieck ring of characters, so it is no surprise that the Verlinde formula reconstructs the structure constants of this ring, rather than that of the full fusion ring.  And as we have seen, these structure constants are quite often negative.

\section*{Acknowledgements}

I would like to thank Vladimir Mitev for initiating these thoughts by asking me what the extended algebra of $\func{\affine{\alg{sl}}}{2}_{-1/2}$ would be.  I also thank Volker Schomerus for introducing me to spectral flow, Thomas Creutzig for enlightening discussions on what this actually means, and Pierre Mathieu for explaining \cite{LesSU202} to me.  This work was partially supported by the Galileo Galilei Institute for Theoretical Physics, the INFN, and the Marie Curie Excellence Grant MEXT-CT-2006-042.

\appendix

\section{Spectral Flow} \label{appSpecFlow}

In this appendix, we detail the construction of spectral flow automorphisms.  Spectral flow has a long history in the \cft{} literature, and can be traced back at least as far as \cite{SchCom87}.  The name refers to the fact that these automorphisms do not preserve the conformal dimension, hence the spectrum ``flows'' (discretely in this case) under their action.  We are actually only interested in the case where $\affine{\alg{g}} = \func{\affine{\alg{sl}}}{2}$, but it is not much harder to develop the theory for general (untwisted) affine Kac-moody algebras $\affine{\alg{g}}$ (and it is very beautiful).

\subsection{Affine Weyl Group Translations} \label{appAffW}

Let $\alg{g}$ be the horizontal subalgebra of $\affine{\alg{g}}$, let $\alpha$ denote a root of $\alg{g}$ with root vector $e^{\alpha}$ and coroot $\alpha^{\vee}$, and let $\group{W}$ be the Weyl group of $\alg{g}$.  Then, each $w \in \group{W}$ permutes the roots and thereby induces an automorphism of $\alg{g}$ via
\begin{equation}
\func{w}{e^{\alpha}} = e^{\func{w}{\alpha}}, \qquad \text{hence} \qquad \func{w}{\alpha^{\vee}} = \func{w}{\alpha}^{\vee}.
\end{equation}
This generalises to $\affine{\alg{g}}$ as follows.  The real roots now take the form $\alpha + n \affine{\delta}$ ($n \in \ZZ$), where $\alpha$ is a root of $\alg{g}$ and $\affine{\delta}$ is the generating imaginary root.  The corresponding root vector is $e^{\alpha}_n$.  The root vectors corresponding to the imaginary root $n \affine{\delta}$ ($n \neq 0$) are denoted by $h^i_n$, $i = 1, 2, \ldots, \rank \alg{g}$, and we will associate the $h^i$ with the simple coroots of $\alg{g}$:  $h^i = \alpha_i^{\vee}$.  The affine Weyl group decomposes as $\affine{\group{W}} = \group{W} \ltimes \group{Q}^{\vee}$, where $\group{Q}^{\vee}$ is the coroot lattice of $\alg{g}$.  The coroot lattice acts on the roots of $\affine{\alg{g}}$ by translations in the imaginary direction:
\begin{equation}
\alpha^{\vee} \colon \beta + n \affine{\delta} \longmapsto \beta + \brac{n - \inner{\beta}{\alpha^{\vee}}} \affine{\delta}.
\end{equation}
This is nothing but the usual affine Weyl group action obtained by embedding the roots into the weight space of $\affine{\alg{g}}$.

It follows that the simple coroots $\alpha_i^{\vee}$ ($i=1, 2, \ldots, \rank \alg{g}$) of $\alg{g}$ each define an independent transformation $\tau_i$ on the root vectors of $\affine{\alg{g}}$ via
\begin{equation} \label{eqnStartingPoint}
\func{\tau_i}{e^{\alpha}_n} = e^{\alpha}_{n - \inner{\alpha}{\alpha_i^{\vee}}} \quad \text{($n \in \ZZ$)} \qquad \text{and} \qquad \func{\tau_i}{h^j_n} = h^j_n \quad \text{($n \neq 0$).}
\end{equation}
We extend these transformations to automorphisms of $\affine{\alg{g}}$.  First we compute
\begin{align}
\func{\tau_i}{h^j_0} &= \func{\tau_i}{\comm{e^{\alpha_j}_n}{e^{-\alpha_j}_{-n}} - n \killing{e^{\alpha_j}}{e^{-\alpha_j}} K} = \comm{e^{\alpha_j}_{n - \inner{\alpha_j}{\alpha_i^{\vee}}}}{e^{-\alpha_j}_{-n + \inner{\alpha_j}{\alpha_i^{\vee}}}} - \frac{2n}{\norm{\alpha_j}^2} \func{\tau_i}{K} \notag \\
&= h^j_0 - \frac{2 \inner{\alpha_j}{\alpha_i^{\vee}}}{\norm{\alpha_j}^2} K + \frac{2n}{\norm{\alpha_j}^2} \brac{K - \func{\tau_i}{K}}.
\end{align}
Here, $\killing{\cdot}{\cdot}$ denotes the Killing form of $\alg{g}$.  Since this computation holds for all $n \in \ZZ$, we must have
\begin{equation}
\func{\tau_i}{h^j_0} = h^j_0 - \killing{\alpha_i^{\vee}}{\alpha_j^{\vee}} K \qquad \text{and} \qquad \func{\tau_i}{K} = K.
\end{equation}

It remains to determine the action of the $\tau_i$ on $L_0$.  This is fixed by the Sugawara construction, but requires a little work.  The normal-ordering appearing in this construction turns out to cause some difficulties and we will treat these by working in the (equivalent) field-theoretic framework, rather than at the level of the algebra itself.  Note that the automorphisms $\tau_i$ act on the fields $\func{e^{\alpha}}{z} = \sum_n e^{\alpha}_n z^{-n-1}$ and $\func{h^j}{z} = \sum_n h^j_n z^{-n-1}$ by
\begin{equation}
\func{\tau_i}{\func{e^{\alpha}}{z}} = z^{-\inner{\alpha}{\alpha_i^{\vee}}} \func{e^{\alpha}}{z} \qquad \text{and} \qquad \func{\tau_i}{\func{h^j}{z}} = \func{h^j}{z} - \killing{\alpha_i^{\vee}}{\alpha_j^{\vee}} K z^{-1}.
\end{equation}
Our goal is therefore to determine the corresponding action on
\begin{equation} \label{eqnDefT}
\func{T}{z} = \frac{1}{2 \brac{K + \dCox}} \sqbrac{\sum_{m,n=1}^{\rank \alg{g}} \invkilling{h^m}{h^n} \normord{\func{h^m}{z} \func{h^n}{z}} + \sum_{\alpha \in \Delta} \invkilling{e^{\alpha}}{e^{-\alpha}} \normord{\func{e^{\alpha}}{z} \func{e^{-\alpha}}{z}}},
\end{equation}
where $\dCox$ is the dual Coxeter number of $\alg{g}$ and $\Delta$ is the set of roots of $\alg{g}$.

We first note that
\begin{equation}
\func{\tau_i}{\normord{\func{h^m}{z} \func{h^n}{z}}} = \normord{\func{h^m}{z} \func{h^n}{z}} - \kappa_{im} \func{h^n}{z} K z^{-1} - \kappa_{in} \func{h^m}{z} K z^{-1} + \kappa_{im} \kappa_{in} K^2 z^{-2},
\end{equation}
where $\kappa_{ab} = \kappa_{ba}$ denotes $\killing{h^a}{h^b}$.  Under $\tau_i$, the sum over $m$ and $n$ in \eqnref{eqnDefT} therefore gives
\begin{equation}
\sum_{m,n=1}^{\rank \alg{g}} \invkilling{h^m}{h^n} \normord{\func{h^m}{z} \func{h^n}{z}} - 2 \func{h^i}{z} K z^{-1} + \frac{4}{\norm{\alpha_i}^2} K^2 z^{-2}.
\end{equation}
Since $\tau_i$ changes the dimension of the $\func{e^{\alpha}}{z}$, it affects the normal-ordering in the corresponding terms in a non-trivial way.  Using the standard definition of normal-ordering in \cft{}, we compute
\begin{align}
\func{\tau_i}{\normord{\func{e^{\alpha}}{w} \func{e^{-\alpha}}{w}}} &= \oint_w \func{e^{\alpha}}{z} \func{e^{-\alpha}}{w} z^{-\inner{\alpha}{\alpha_i^{\vee}}} w^{\inner{\alpha}{\alpha_i^{\vee}}} \brac{z-w}^{-1} \frac{\dd z}{2 \pi \ii} \notag \\
&= w^{\inner{\alpha}{\alpha_i^{\vee}}} \oint_w z^{-\inner{\alpha}{\alpha_i^{\vee}}} \sqbrac{\frac{2K / \norm{\alpha}^2}{\brac{z-w}^3} + \frac{\func{\alpha^{\vee}}{w}}{\brac{z-w}^2} + \frac{\normord{\func{e^{\alpha}}{w} \func{e^{-\alpha}}{w}}}{z-w}}  \frac{\dd z}{2 \pi \ii} \notag \\
&= \normord{\func{e^{\alpha}}{w} \func{e^{-\alpha}}{w}} - \inner{\alpha}{\alpha_i^{\vee}} w^{-1} \func{\alpha^{\vee}}{w} + \frac{\inner{\alpha}{\alpha_i^{\vee}} \brac{\inner{\alpha}{\alpha_i^{\vee}} + 1}}{\norm{\alpha}^2} K w^{-2}.
\end{align}
Under $\tau_i$, the sum over the roots in \eqnref{eqnDefT} gives
\begin{align}
\sum_{\alpha \in \Delta} &\sqbrac{\invkilling{e^{\alpha}}{e^{-\alpha}} \normord{\func{e^{\alpha}}{z} \func{e^{-\alpha}}{z}} - \frac{\norm{\alpha}^2}{2} \inner{\alpha}{\alpha_i^{\vee}} z^{-1} \func{\alpha^{\vee}}{z} + \frac{\inner{\alpha}{\alpha_i^{\vee}} \brac{\inner{\alpha}{\alpha_i^{\vee}} + 1}}{2} K z^{-2}} \notag \\
&= \sum_{\alpha \in \Delta} \sqbrac{\invkilling{e^{\alpha}}{e^{-\alpha}} \normord{\func{e^{\alpha}}{z} \func{e^{-\alpha}}{z}} - \frac{2}{\norm{\alpha_i}^2} \brac{\alpha , \alpha_i} z^{-1} \func{\alpha}{z} + \frac{2}{\norm{\alpha_i}^4} \brac{\alpha_i , \alpha} \brac{\alpha , \alpha_i} K z^{-2}} \notag \\
&= \sum_{\alpha \in \Delta} \invkilling{e^{\alpha}}{e^{-\alpha}} \normord{\func{e^{\alpha}}{z} \func{e^{-\alpha}}{z}} - 2 \dCox z^{-1} \func{\alpha_i^{\vee}}{z} + \frac{4 \dCox}{\norm{\alpha_i}^2} K z^{-2}.
\end{align}
Here in the first step, we have used the fact that summands over $\Delta$ which are odd under $\alpha \rightarrow - \alpha$ give vanishing sums.  In the second step, we use (twice) the fact that
\begin{equation}
\sum_{\alpha \in \Delta} \brac{\lambda , \alpha} \brac{\alpha , \mu} = 2 \dCox \brac{\lambda , \mu}
\end{equation}
for all weights $\lambda$ and $\mu$.

Putting this all together (and remembering that $h^i = \alpha_i^{\vee}$), we finally obtain
\begin{gather}
\func{\tau_i}{\func{T}{z}} = \func{T}{z} - z^{-1} \func{h^i}{z} + \frac{2}{\norm{\alpha_i}^2} K z^{-2} \\
\Rightarrow \qquad \func{\tau_i}{L_0} = L_0 - h^i_0 + \frac{2}{\norm{\alpha_i}^2} K.
\end{gather}
This then completes the description of the automorphisms of $\affine{\alg{g}}$ induced by the translation subgroup of the affine Weyl group.  It is not hard to check now that powers of $\tau_i$ act as follows:
\begin{subequations} \label{eqnTransAuts}
\begin{gather}
\func{\tau_i^{\ell}}{e^{\alpha}_n} = e^{\alpha}_{n - \ell \inner{\alpha}{\alpha_i^{\vee}}} \qquad \func{\tau_i^{\ell}}{h^j_n} = h^j_n - \ell \killing{\alpha_i^{\vee}}{\alpha_j^{\vee}} \delta_{n,0} K \\
\func{\tau_i^{\ell}}{K} = K \qquad \func{\tau_i^{\ell}}{L_0} = L_0 - \ell h^i_0 + \ell \brac{\ell + \frac{2}{\norm{\alpha_i}^2} - 1} K.
\end{gather}
\end{subequations}
These automorphisms are examples of \emph{spectral flow} automorphisms.  However, they do not usually exhaust the latter in general, as we shall see.

\subsection{Outer Automorphisms} \label{appAffOut}

Having determined the explicit action of the algebra automorphisms induced by the affine Weyl group, we can turn to the remaining automorphisms of $\affine{\alg{g}}$, the outer automorphisms induced by the symmetries of the Dynkin diagram.  Unlike the (non-trivial) affine Weyl transformations, these preserve a given set of a simple roots.  Indeed, an outer automorphism is completely determined by the permutation it induces on the (chosen set of) simple roots.

The outer automorphisms of $\alg{g}$ therefore just permute the root vectors $e^{\alpha}_n$ and $h^j_n$ of $\affine{\alg{g}}$ without changing the grade $n$.  But, by analogy with the results of the previous section, we would like to understand the general case.  Happily, this is a simple endeavour.  The automorphisms of $\affine{\alg{g}}$ which preserve the chosen Cartan subalgebra can be decomposed into
\begin{equation}
\Aut \affine{\alg{g}} = \Out \affine{\alg{g}} \ltimes \affine{\group{W}} = \Aut \alg{g} \ltimes \group{Q}^*,
\end{equation}
where $\group{Q}^*$ denotes the dual of the root lattice.  Thus, our endeavour corresponds to generalising the results of \appref{appAffW} to the outer automorphisms of $\alg{g}$ (which is trivial) and replacing coroot lattice translations by dual root lattice translations.  It is these dual root translations which generate the complete set of spectral flow automorphisms.

In fact, it is easy to understand these latter translations.  Recall from \eqnref{eqnStartingPoint} that our starting point for constructing the automorphisms corresponding to a translation by the simple coroot $\alpha_i^{\vee}$ was the effect on $e^{\alpha}_n$.  Everything else follows from this effect, which was to lower $n$ by $\inner{\alpha}{\alpha_i^{\vee}}$.  However, this index will still be an integer (for all roots $\alpha$) if we replace $\alpha_i^{\vee}$ by an element of the dual root lattice $\group{Q}^*$, so it follows that such a replacement will still lead to a well-defined automorphism of $\affine{\alg{g}}$.

In fact, we can always choose a basis of $\group{Q}^*$ whose $\rank \alg{g}$ elements are of the form $q_i^{\vee} / m_i$ for some $q_i^{\vee} \in \group{Q}^{\vee}$ and $m_i \in \ZZ$ (the fact that $\group{Q}^*$ contains $\group{Q}^{\vee}$ follows from the integrality of the Cartan matrix).  We may therefore determine generators of the automorphism group corresponding to dual root lattice translations by finding such a basis and applying \eqnref{eqnTransAuts} with $\ell$ fractional.  Note however that scaling $\alpha^{\vee}$ by some factor $t$ corresponds to scaling $\alpha$ by $t^{-1}$.

For example, the coroot lattice of $\func{\affine{\alg{sl}}}{2}$ is generated by $\alpha_1^{\vee}$, so the coroot spectral flow automorphisms are generated by $\tau_1$:
\begin{equation}
\func{\tau_1}{e^{\alpha}_n} = e^{\alpha}_{n - 2}, \qquad \func{\tau_1}{h^1_n} = h^1_n - 2 \delta_{n,0} K, \qquad \func{\tau_1}{K} = K, \qquad \func{\tau_1}{L_0} = L_0 - h^1_0 + K.
\end{equation}
The dual root lattice is however generated by $\alpha_1^{\vee} / 2$.  It follows that the spectral flow automorphisms are generated by $\gamma = \tau_1^{1/2}$.  By \eqnref{eqnTransAuts}, the action of $\gamma$ is given by
\begin{equation} \label{eqnSpecFlowA1}
\func{\gamma}{e^{\alpha}_n} = e^{\alpha}_{n - 1}, \qquad \func{\gamma}{h^1_n} = h^1_n - \delta_{n,0} K, \qquad \func{\gamma}{K} = K, \qquad \func{\gamma}{L_0} = L_0 - \frac{1}{2} h^i_0 + \frac{1}{4} K.
\end{equation}
It should be clear from these formulae why $\tau_1$ has a square root.

As a second example, the dual root lattice of $\func{\affine{\alg{sl}}}{3}$ is generated by $\tfrac{2}{3} \alpha_1^{\vee} + \tfrac{1}{3} \alpha_2^{\vee}$ and $\tfrac{1}{3} \alpha_1^{\vee} + \tfrac{2}{3} \alpha_2^{\vee}$.  We therefore have the spectral flow generators $\gamma_1 = \tau_1^{2/3} \tau_2^{1/3}$ and $\gamma_2 = \tau_1^{1/3} \tau_2^{2/3}$, which act on $\func{\affine{\alg{sl}}}{3}$ via
\begin{subequations}
\begin{gather}
\func{\gamma_i}{e^{\alpha_j}_n} = e^{\alpha_j}_{n - \delta_{i,j}}, \qquad \func{\gamma_i}{e^{\theta}_n} = e^{\theta}_{n - 2}, \qquad \func{\gamma_i}{h^j_n} = h^j_n - \delta_{n,0} \delta_{i,j} K, \\
\func{\gamma_i}{K} = K, \qquad \func{\gamma_i}{L_0} = L_0 - \frac{1}{3} \brac{h^1_0 + h^2_0} - \frac{1}{3} h^i_0 + \frac{1}{3} K.
\end{gather}
\end{subequations}

Finally, note that composing any representation of $\affine{\alg{g}}$ with an automorphism gives another representation.  Hence, spectral flow automorphisms induce maps (vector space isomorphisms) between $\affine{\alg{g}}$-modules.  Since such maps must preserve integrability, the set of integrable $\affine{\alg{g}}$-modules must close under the induced spectral flow.  In fact, integrable modules are mapped to themselves when the spectral flow corresponds to a translation by a coroot lattice element.  More general translations induce maps between integrable modules whose highest weights are related by an outer automorphism.  In both cases, these maps are non-trivial and provide a wealth of information about the integrable modules.  When the modules are not integrable, the spectral flow generally does not map any module to itself, even if the flow corresponds to a coroot translation.  In this case, spectral flow automorphisms are useful for understanding the spectrum and for investigating the structure of the unfamiliar modules which arise.

\section{Jacobi Theta Functions} \label{appTheta}

We collect here for convenience our notation for the Jacobi theta functions and some of their important properties.  First we define
\begin{subequations}
\begin{align}
\Jth{1}{z ; q} &= -\ii \sum_{n \in \ZZ} \brac{-1}^n z^{n+1/2} q^{\brac{n+1/2}^2 / 2} & \Jth{3}{z ; q} &= \sum_{n \in \ZZ} z^n q^{n^2 / 2} \\
\Jth{2}{z ; q} &= \sum_{n \in \ZZ} z^{n+1/2} q^{\brac{n+1/2}^2 / 2} & \Jth{4}{z ; q} &= \sum_{n \in \ZZ} \brac{-1}^n z^n q^{n^2 / 2}.
\end{align}
\end{subequations}
From these definitions follow a number of simple relations:
\begin{subequations} \label{eqnThId-z}
\begin{align}
\Jth{1}{e^{\ii \pi} z ; q} &= \Jth{2}{z ; q} & \Jth{3}{e^{\ii \pi} z ; q} &= \Jth{4}{z ; q} \\
\Jth{2}{e^{\ii \pi} z ; q} &= -\Jth{1}{z ; q} & \Jth{4}{e^{\ii \pi} z ; q} &= \Jth{3}{z ; q}
\end{align}
\end{subequations}
\begin{subequations} \label{eqnThIdSF}
\begin{align}
\Jth{1}{zq^{1/2} ; q} &= \frac{\ii}{z^{1/2} q^{1/8}} \Jth{4}{z ; q} & \Jth{3}{zq^{1/2} ; q} &= \frac{1}{z^{1/2} q^{1/8}} \Jth{2}{z ; q} \\
\Jth{2}{zq^{1/2} ; q} &= \frac{1}{z^{1/2} q^{1/8}} \Jth{3}{z ; q} & \Jth{4}{zq^{1/2} ; q} &= \frac{\ii}{z^{1/2} q^{1/8}} \Jth{1}{z ; q}
\end{align}
\end{subequations}
By making use of Jacobi's triple product identity \cite[Eq.~2.2.10]{AndThe76},
\begin{equation} \label{eqnJacobiTriple}
\prod_{i=1}^{\infty} \brac{1 + z q^{i-1/2}} \brac{1 - q^i} \brac{1 + z^{-1} q^{i-1/2}} = \sum_{n \in \ZZ} z^n q^{n^2 / 2},
\end{equation}
each of the theta functions may be written in product form:
\begin{subequations} \label{eqnThIdPF}
\begin{align}
\Jth{1}{z ; q} &= -\ii z^{1/2} q^{1/8} \prod_{i=1}^{\infty} \brac{1 - z q^i} \brac{1 - q^i} \brac{1 - z^{-1} q^{i-1}} \\
\Jth{2}{z ; q} &= z^{1/2} q^{1/8} \prod_{i=1}^{\infty} \brac{1 + z q^i} \brac{1 - q^i} \brac{1 + z^{-1} q^{i-1}} \\
\Jth{3}{z ; q} &= \prod_{i=1}^{\infty} \brac{1 + z q^{i-1/2}} \brac{1 - q^i} \brac{1 + z^{-1} q^{i-1/2}} \\
\Jth{4}{z ; q} &= \prod_{i=1}^{\infty} \brac{1 - z q^{i-1/2}} \brac{1 - q^i} \brac{1 - z^{-1} q^{i-1/2}}.
\end{align}
\end{subequations}
This also gives us the identity
\begin{equation} \label{eqnIdeta}
\Jth{2}{1 ; q} \Jth{3}{1 ; q} \Jth{4}{1 ; q} = 2 \func{\eta}{q}^3,
\end{equation}
where $\eta$ is Dedekind's eta function
\begin{equation} \label{eqnDefeta}
\func{\eta}{q} = q^{1/24} \prod_{i=1}^{\infty} \brac{1 - q^i}.
\end{equation}

The most important property of these functions is their behaviour under modular transformations.  Setting $z = \func{\exp}{2 \pi \ii \zeta}$ and $q = \func{\exp}{2 \pi \ii \tau}$, the modular group $\func{\group{SL}}{2 ; \ZZ}$ is generated by two transformations $S$ and $T$ which act via
\begin{equation} \label{eqnModGenAct}
S \colon \brac{\zeta , \tau} \longmapsto \brac{\zeta / \tau , -1 / \tau} \qquad T \colon \brac{\zeta , \tau} \longmapsto \brac{\zeta , \tau + 1}.
\end{equation}
One can check that $S^4 = \brac{ST}^6 = \id$.  Writing $\Jth{i}{\zeta \mid \tau}$ for $\Jth{i}{e^{2 \pi \ii \zeta} ; e^{2 \pi \ii \tau}}$, $T$ is therefore represented on the space of theta functions by
\begin{subequations} \label{eqnThIdT}
\begin{align}
\Jth{1}{\zeta \mid \tau + 1} &= e^{\ii \pi / 4} \Jth{1}{\zeta \mid \tau} & \Jth{3}{\zeta \mid \tau + 1} &= \Jth{4}{\zeta \mid \tau} \\
\Jth{2}{\zeta \mid \tau + 1} &= e^{\ii \pi / 4} \Jth{2}{\zeta \mid \tau} & \Jth{4}{\zeta \mid \tau + 1} &= \Jth{3}{\zeta \mid \tau}.
\end{align}
\end{subequations}
\eqnref{eqnDefeta} gives (in hopefully obvious notation)
\begin{equation} \label{eqnetaT}
\func{\eta}{\tau + 1} = e^{\ii \pi / 12} \func{\eta}{\tau}.
\end{equation}
Determining the corresponding transformations under $S$ requires a specialisation of the Poisson resummation formula from Fourier analysis.  With this tool, we derive
\begin{subequations} \label{eqnThIdS}
\begin{align}
\Jth{1}{\zeta / \tau , -1 / \tau} &= -\ii \sqrt{-\ii \tau} \ e^{\ii \pi \zeta^2 / \tau} \Jth{1}{\zeta \mid \tau} & \Jth{3}{\zeta / \tau , -1 / \tau} &= \sqrt{-\ii \tau} \ e^{\ii \pi \zeta^2 / \tau} \Jth{3}{\zeta \mid \tau} \\
\Jth{2}{\zeta / \tau , -1 / \tau} &= \sqrt{-\ii \tau} \ e^{\ii \pi \zeta^2 / \tau} \Jth{4}{\zeta \mid \tau} & \Jth{4}{\zeta / \tau , -1 / \tau} &= \sqrt{-\ii \tau} \ e^{\ii \pi \zeta^2 / \tau} \Jth{2}{\zeta \mid \tau}.
\end{align}
\end{subequations}
The additional factor of $-\ii$ for $\vartheta_1$ reflects the fact that this theta function is antisymmetric under $z \rightarrow z^{-1}$ whereas the others are symmetric.  \eqnref{eqnIdeta} now gives
\begin{equation} \label{eqnetaS}
\func{\eta}{-1 / \tau} = \sqrt{-\ii \tau} \ \func{\eta}{\tau}.
\end{equation}

\end{document}